\documentclass[12pt,twocolumn,tighten]{aastex62}
\usepackage{hyperref}
\usepackage{color}
\usepackage{natbib}
\usepackage{rotating}
\usepackage{amsmath,tabu}
\usepackage{threeparttable}
\usepackage{eqnarray}

\begin{document}

\title{Exploring the Physical Properties of
the Cool Circumgalactic Medium with a Semi-Analytic Model}

\author{Ting-Wen Lan}
\affil{Kavli IPMU, the University of Tokyo (WPI), Kashiwa 277-8583, Japan}
\author{Houjun Mo}
\affil{Department of Astronomy, University of Massachusetts, LGRT-B619E, 710 North Pleasant Street, Amherst, MA, 01003, USA}
\affil{Physics Department and Center for Astrophysics, Tsinghua University, Beijing 10084, China}

\begin{abstract}
 We develop a semi-analytic model to explore the physical properties of cool pressure-confined
 circumgalactic clouds with mass ranging from $10$ to $10^{8} \, \rm M_{\odot}$ in a hot diffuse halo. 
 We consider physical effects that control the motion and mass loss of the clouds, and estimate the lifetime 
 and the observed properties of individual cool gas clouds inferred from the CLOUDY simulation. Our results show that the cool pressure-confined gas clouds have physical properties consistent
 with absorption line systems with neutral hydrogen column densities $N_{\rm HI}\geq10^{18.5} \rm cm^{-2}$ such as strong metal absorbers, sub-DLAs, and DLAs. The cool circumgalactic clouds are transient due to evaporation and recycling and 
 therefore a constant replenishment is needed to maintain the cool CGM.
 We further model the ensemble properties of the cool CGM with clouds originated from outflows, inflows, 
 or/and in-situ formation with a range of initial cloud mass function and velocity distribution.
 We find that only with a certain combination of parameters, an outflow model can broadly reproduce three cool gas properties 
 around star-forming galaxies simultaneously: the spatial distribution, down-the-barrel outflow absorption, 
 and gas velocity dispersion. Both a constant insitu model and gas inflow model can 
 reproduce the observed covering fractions of high $\rm N_{HI}$ gas around passive galaxies 
 but they fail to reproduce sufficient number of low $\rm N_{HI}$ systems. 
 The limitations and the failures of the current models are discussed. 
 Our results illustrate that semi-analytic modeling is a promising tool to understand 
 the physics of the cool CGM which is usually unresolved by state-of-the-art 
 cosmological hydrodynamic simulations.
\end{abstract}

%%%%%%%%%%%%%%% SECTION 1 %%%%%%%%%%%%%%%%%%%%%%
\section{Introduction}

Gas around galaxies, the cirumgalactic medium \citep[CGM,][for a review]{Tumlinson2017}, plays 
a key role in galaxy formation and evolution. As the interface between galaxies and the intergalactic medium (IGM), the CGM contains signatures of gas flow processes that drive the evolution 
of galaxies. Since the first discovery of the CGM \citep{Bergeron1986}, studies have revealed the 
complex nature of the CGM. The CGM is multi-phase, consisting of gas with wide ranges of temperature and density.
In addition, the properties of galaxies 
and their CGM have complicated relationships; while cool gas ($\sim10^{4}$ K) appears to exist around 
both star-forming and passive galaxies \citep[e.g.,][]{Chen2010, Thom2012, Lan2014}, warm gas is much more abundant around star-forming galaxies than 
around passive galaxies \citep[e.g.,][]{Tumlinson2011}. 

To interpret these observational results, simulations and theories for the CGM have been developed. 
Recently, many studies have attempted to address the origin of the excess of warm gas traced by OVI 
absorption lines (diffuse gas with $\sim10^{5.5}$ K assuming collisional ionization) around star-forming galaxies 
\citep[e.g.,][]{Bordoloi2017, Faerman2017,math2017, Nelson2018, McQuinn2018, Stern2018}. On the other hand, 
although cool gas has been observed ubiquitously around galaxies (as, e.g., traced by MgII absorption lines), 
only limited attempts have been made to model and understand the origin of the cool halo gas 
\citep[e.g.,][]{Mo1996, MB2004, Kaufmann2009, Bordoloi2014, Faucher2016, vandevoort2018}. 
More importantly, observations \citep[e.g.,][]{Rauch1999, Rigby2002, p2009, Lan2017} and 
simulations \citep[e.g.,][]{McCourt2016, Liang2018, Sparre2018, Surech2018} have shown that the physical 
scale relevant to the cool circumgalactic gas is of the order of a few tens of parsec or even smaller, which is 
several orders of magnitude smaller than the resolution of any current galaxy hydrodynamic 
simulations \citep[see Figure 9 in][]{Sparre2018}. 
This discrepancy poses a challenge to understanding the physics for the CGM and 
galaxy formation. 

To overcome the limitation of numerical resolution, we develop a semi-analytic model for the cool 
circumgalactic gas, motivated by earlier analytic works. We revisit the idea of cool gas clouds being in 
pressure equilibrium with an ambient hot halo gas \citep{Mo1996, MB2004} and estimate their lifetime, motion, 
and trajectory by taking into account the effects of gravity, ram pressure and heat evaporation. 
We model the ensemble properties of the cool gas clouds in halos considering scenarios of 
outflows, inflows, and in-situ formation. With this flexible model, our goal is to explore how physical mechanisms affect the properties of the CGM and to constrain parameter space of these mechanisms with observations.

This paper is organized as follows. In Section 2, we describe the considered physical effects 
for cool gas clouds. In Section 3, we summarize the 
evolution of individual clouds and present the ensemble properties of cool gas and 
their evolution. We discuss other implications in Section 4 and summarize our results in Section 5. 
Throughout the paper we adopt a flat $\Lambda$CDM cosmology with $h=0.7$ and 
$\Omega_{\rm M}=0.3$.

%======================================================
%Figure 
%======================================================
\begin{figure}[]
\center
\includegraphics[scale=0.4]{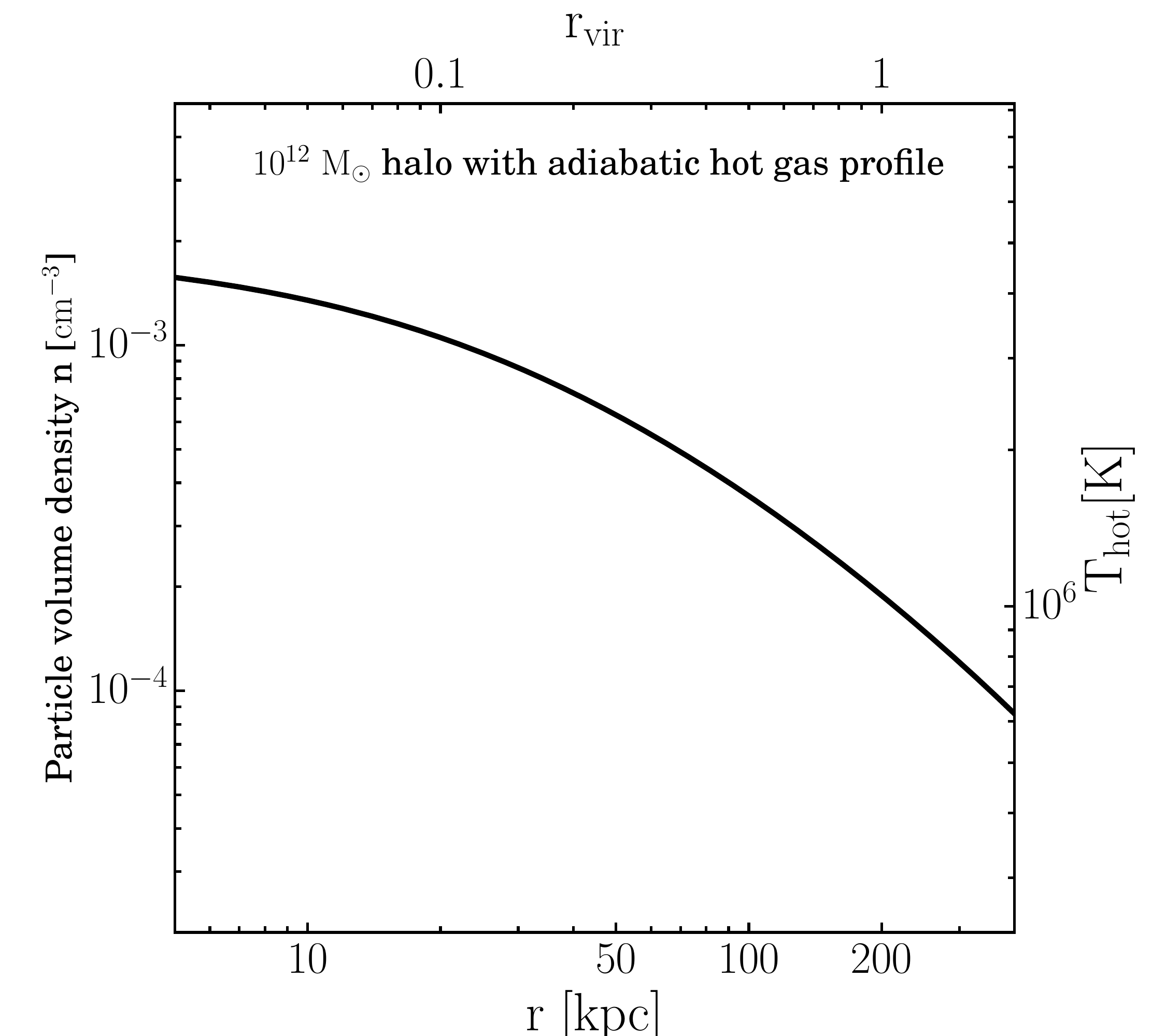}
\caption{Hot gas density and temperature profiles.}
\label{fig:hot_gas_profiles}
\end{figure}
%======================================================

%
%======================================================
%Figure 
%======================================================
\begin{figure}[]
\center
\includegraphics[scale=0.38]{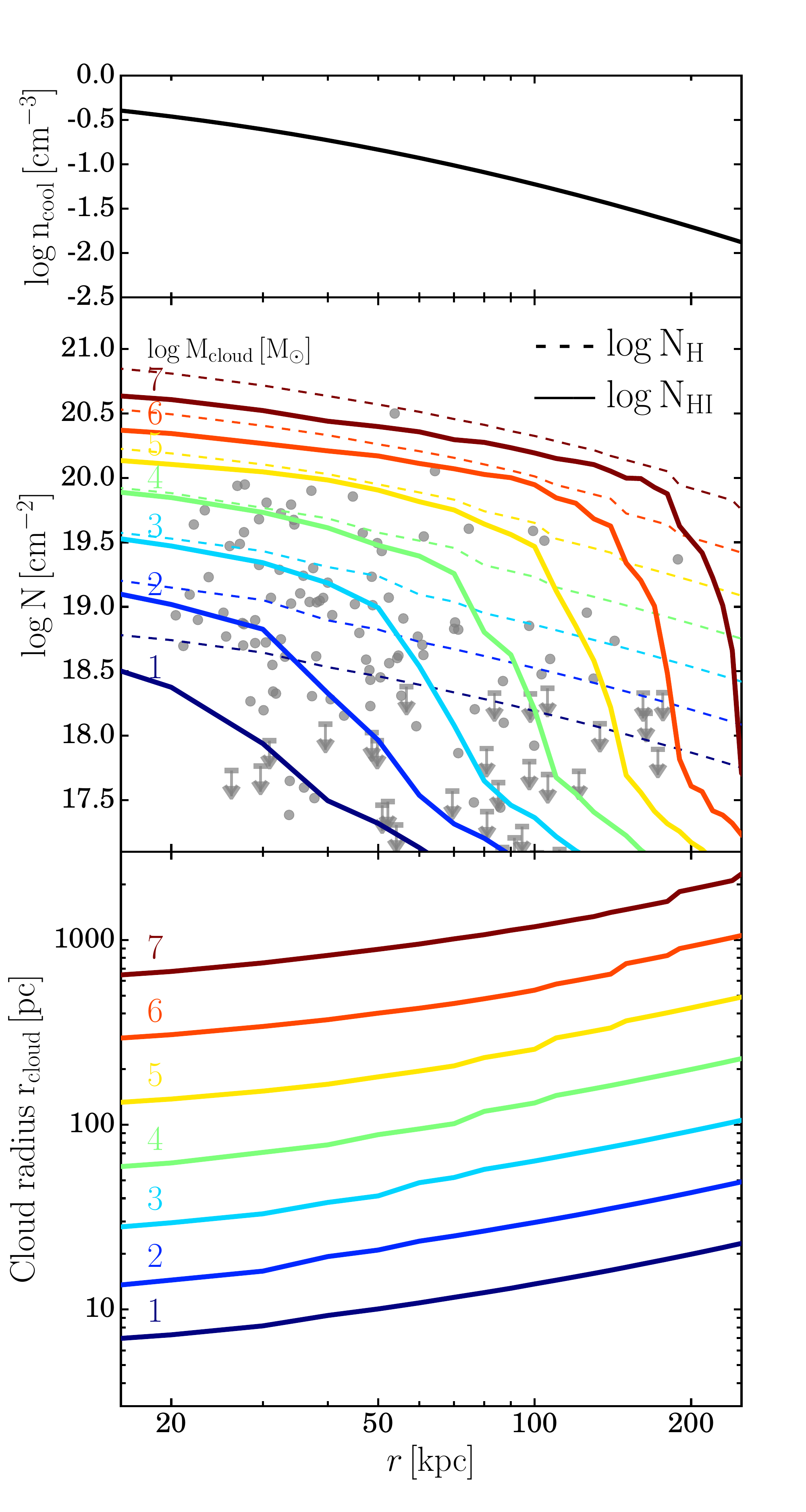}
\caption{Cloud properties as a function of position. {\it Top:} $10^{4}$ K cool gas volume density.
{\it Middle:} neutral hydrogen  (solid color lines) and total hydrogen  (dashed color lines) column 
densities as a function of cloud mass. The data points show the neutral hydrogen column densities traced by 
strong MgII absorbers from observations. Arrows indicate upper limits.
Note that data points are in the projected distance. 
{\it Bottom:} cloud radius as a function of mass.}
\label{fig:hot_gas_profiles}
\end{figure}
%======================================================
%

%%%%%%%%% SECTION 2 %%%%%%%%%%%%%%%%%%%%%%%%%
\section{Model setup}

To explore the properties of cool pressure-confined clouds in the CGM,  
we first introduce the model components that control the motion and the lifetime of 
individual clouds (\S2.1). We then describe our outflow and inflow models 
to populate an ensemble of clouds into the CGM and to characterize the 
statistical properties of the cool clouds (\S2.2). 

\subsection{Model components for individual clouds}

\textbf{Basic properties-} The evolution of cool clouds is modeled under the 
influence of a dark matter halo and a hot diffuse gas halo. We adopt a $10^{12}\ \rm M_{\odot}$ dark matter halo, which corresponds to $\sim \rm L^*$ galaxies \citep{Yang2003,Yang2012}, with a 
NFW profile \citep{NFW1996} and concentration $c=10$. 
This halo mass is chosen in order to compare with available observational measurements. The effect of halo mass will be discussed in Section 4.
For the hot gaseous 
halo, we use the gas profile derived by \citet{MB2004}, assuming that the initial 
hot gas had cooled over 6 billion years and the residual hot gas had reached 
hydrostatic equilibrium adiabatically at $z=0.5$. The metallicity of the hot gas 
is assumed to be 0.1 solar. Figure 1 shows the adopted hot gas density 
(left axis) and the temperature (right axis) profiles. 
This hot gas profile is similar to the observed hot gas profile of the Milky Way \citep[e.g.,][]{Gupta2017}.

The physical properties of cool gas clouds are obtained by assuming 
that \textit{cool gas clouds are in pressure equilibrium with the 
surrounding hot gas halo} \citep[e.g.][]{Mo1996},
\begin{equation}
  n_{\rm cloud}\, T_{\rm cloud} =n_{\rm hot}\, T_{\rm hot}.
\end{equation}
This is motivated by the fact that a gas cloud can establish pressure equilibrium 
with its surrounding typically within the sound crossing time. 
For a cloud with $10^{4}\,\rm M_{\odot}$, $n\sim 0.3 \ \rm cm^{-3}$, 
and $T\sim10^{4}\,K$, the sound crossing time is about 10 Myr, 
much shorter than the survival time of the clouds, which is greater than 100 Myr. 
We assume that the temperature of gas clouds is $10^{4}\,\rm K$ with 
spherical geometry. The physical properties of a cool gas cloud 
of a given mass $M_{\rm cloud}$, such as its size $r_{\rm cloud}$, 
volume density $n_{\rm cloud}$, the total hydrogen column density 
$N_{\rm H}$, neutral hydrogen column density $N_{\rm HI}$, and 
ionization fraction, are then obtained self-consistently by 
using CLOUDY simulation \citep{CLOUDY2013}.  Gas clouds are 
assumed to be photo-ionized by an extra-galactic radiation field 
described by the model of \citet{Haardt2001}\footnote{We use the unpublished 
updated 2005 version implemented in the CLOUDY simulation.}. 

In Figure 2, we show the properties of cool clouds as functions of cloud mass 
and location in the halo. The top panel shows the volume densities of the cool gas clouds. 
In the middle panel, the solid color lines show the neutral hydrogen column 
densities $N_{\rm HI}$ of clouds and the dashed lines show the total hydrogen 
column densities. The neutral hydrogen column densities fall within the range 
between $10^{18}$ and $10^{20.5}\rm \ cm^{-2}$, similar to that observed in 
sub-damped Lyman systems (strong Lyman-limited systems) and damped Lyman 
systems. For comparison, the grey squares show the neutral hydrogen column 
densities traced by strong MgII absorbers, obtained by using MgII 
galaxy-absorber pairs from \citet{Nielsen2013} and the empirical relationship 
between MgII rest equivalent width and neutral hydrogen column density
from \citet{Lan2017}:
\begin{equation}
    \rm N_{HI}\simeq10^{19}\bigg(\frac{W_{\lambda2796}}{1 \AA}\bigg)^{1.7}\bigg(1+z\bigg)^{1.9} cm^{-2}.
\end{equation} 
Note, however,  that the grey data points use the projected distance 
instead of the three-dimensional distance in the halo.

At small scales, the neutral hydrogen column density depends on the volume 
density of clouds:
\begin{equation}
\rm N_{HI} \propto n_{HI} \times r_{cloud} \propto n_{H}^{2/3}\, M_{cloud}^{1/3}. 
\end{equation}
However, beyond a certain distance scale, the neutral hydrogen column densities drop 
rapidly due to that the volume density of the cloud is too low to be self-shielded 
from the ionization radiation field. This scale depends on the cloud mass, 
as shown in the figure. In the bottom panel of Figure 2, we show cloud radius
(assuming spherical shape) as a function of mass and distance. The cloud 
radius ranges from 10 pc to a few hundred pc,
with a typical value of $\sim 100\,{\rm pc}$ for cloud mass 
$\sim 10^{4} \, \rm M_{\odot}$.

\textbf{Motion and survival-}
To model the motion and survival of the cool clouds in the halo, we consider the 
following effects: 
\begin{itemize}
    \item \textbf{Motions of clouds} - 
    
    the motion of a cloud is governed by the 
    gravitational potential provided by the dark matter halo and the ram pressure of the 
    hot gas halo:
    \begin{eqnarray}\label{eq:motion}
    M_{\rm cloud} \frac{d^{2}{r}}{dt^{2}} = -\frac{GM_{\rm DM}(<r)M_{\rm cloud}}{r^{2}}
    - F_{\rm ram}
    \end{eqnarray}  
where $G$ is the gravitational constant. The ram pressure drag force from the hot gas 
\citep[e.g.,][]{MB2004}, which always decelerates the cloud, is defined as 
\begin{equation}
F_{\rm ram} = \frac{1}{2}C_{d}\, \rho_{\rm hot}(r)\,v_{\rm cloud}(r)^{2} \times \pi r_{\rm cloud}^{2}
\end{equation}
where $\rho_{\rm hot}(r)$ is the mass density of hot gas at position r, $v_{\rm cloud}(r)$ is the radial velocity of the cloud, and $r_{\rm cloud}$ is the radius of the cloud.
Following \citet{MB2004}, we set ram pressure efficiency, $C_{d}$, to be 1 as the default value but will explore the 
effect of changing its value in Section 3.3.1. The cloud deceleration from ram pressure scales with the column density of clouds as $N_{\rm H}^{-1}$.
Thus, the ram pressure effect is smaller for more massive clouds. In this paper, 
we only consider radial motion of clouds. 

    \item \textbf{Cloud disruptions} - there are several mechanisms that could 
    affect the mass of a cloud. \citet[][]{Mo1996} (see also \citet{MB2004}) provides a 
    detail summary about the possible mechanisms. 
    Here we briefly discuss the key mechanisms:
    
    \textbf{(1) Self-gravity} - The upper limit of a cloud mass in our analysis is 
    constrained by the Jeans mass \citep{Jeans1902} above which the cloud collapses due 
    to its self-gravity. The Jeans mass of clouds in our model setup is about $\rm 10^{8} M_{\odot}$.
    
    \textbf{(2) Hydrodynamic instability} - When a dense cloud moves through a hot gas halo, the cloud 
    is subject to the hydrodynamic instability, such as Kelvin-Helmholtz instability and Rayleigh-Taylor 
    instability. The characteristic timescale of the instabilities \citep[e.g.][]{Murray1993} are
    \begin{equation}
        t_{\rm instability} \sim \frac{r_{\rm cloud}}{V_{\rm cloud}}\sqrt{\frac{T_{\rm Hot}}{T_{\rm cloud}}}.
    \end{equation}
    Based on the characteristic timescale, a cool gas cloud can only travel a 
    distance that is about 10 times its radius (assuming $T_{\rm Hot}\sim100 \ T_{\rm 
    cloud}$) before it is destroyed by the instability. However, several studies have 
    shown that other mechanisms can suppress the effects of hydrodynamical 
    instabilities, such as radiative cooling \citep{Vietri1997}, magnetic 
    fields \citep[e.g.][]{McCourt2015}, and the presence of heat conduction 
    layers \citep{Armillotta2017}, and mixing layer between cool and hot gas 
    \citep{Gronke2018}. Motivated by these studies, we assume that the clouds are 
    intact under the effect of the hydrodynamic instability.
    
    \textbf{(3) Heat evaporation} - Surrounded by hot ambient gas, cool gas clouds are 
    subject to mass loss due to the heating by the hot gas. 
    The classic mass loss rate due to the evaporation is derived by \citet{Cowie1977}:
    \begin{eqnarray}\label{eq:mass_loss}
        \dot{M}_{\rm cloud,classic}&=&\frac{dM_{\rm cloud}}{dt}
        \nonumber\\
        &=& 2.75\times10^{4}\ T_{\rm hot}^{5/2}\ r_{\rm cloud} \bigg(\frac{30}{\ln\Lambda}\bigg) \rm g\,s^{-1}
        \nonumber\\
        &=& 0.44 \, \bigg(\frac{T}{10^{6} \rm \, K}\bigg)^{5/2} \bigg(\frac{r_{\rm cloud}}{\rm pc}\bigg) \, \frac{\rm M_{\odot}}{\rm Myr},
    \end{eqnarray}
    where $\ln \Lambda$ is the Coulomb logarithm with value about 30. 
    \citet{Dalton1993} generalize the mass loss rate including the case when the heat 
    conduction is saturated:
    \begin{equation}
        \dot{M}_{\rm cloud} = f \, w \, \dot{M}_{\rm cloud,classic},
        \label{eq:mass_loss_saturation}
    \end{equation}
    where $w$ is the modified factor depending on the saturation parameter,
    \begin{equation}
         \sigma_{0} = \frac{2 \kappa_{\rm hot} \, T_{\rm hot}}{25 \Phi \, \rho_{\rm hot} 
         \, c_{\rm hot}^{3} \, r_{\rm cloud}}
    \end{equation}
    with $\kappa_{\rm hot}=1.84\times10^{-5}\, T_{\rm hot}^{5/2} \,(\ln\Lambda)^{-1}$
    [cgs], $\Phi$ being an efficiency factor of the order of unity, which is set to be 1, and $c_{\rm hot}$  being the sound speed of the hot gas.
    If $\sigma_{0}$ is greater than 1, the saturation will reduce the mass loss 
    rate to $w \, \dot{M}_{\rm cloud,classic}$. If $0.027<\sigma_{0}<1$, the cloud
    undergoes classical evaporation. Finally, we introduce a suppression factor $f$ for 
    the evaporation rate in Eq. 7. We adopt $f=0.05$ to account for the effect of other 
    mechanisms such as magnetic field \citep[e.g.,][]{Ch1998} on the efficient of mass evaporation. We note that some studies have constrained the $f$ value to be about 0.01 or lower \cite[e.g.,][]{Binney1981, Nipoti2004}.
    We discuss the effect of $f$ value in Section 4.

\end{itemize}

%======================================================
%Figure 
%======================================================
\begin{figure*}[]
\center
\includegraphics[width=0.85\textwidth]{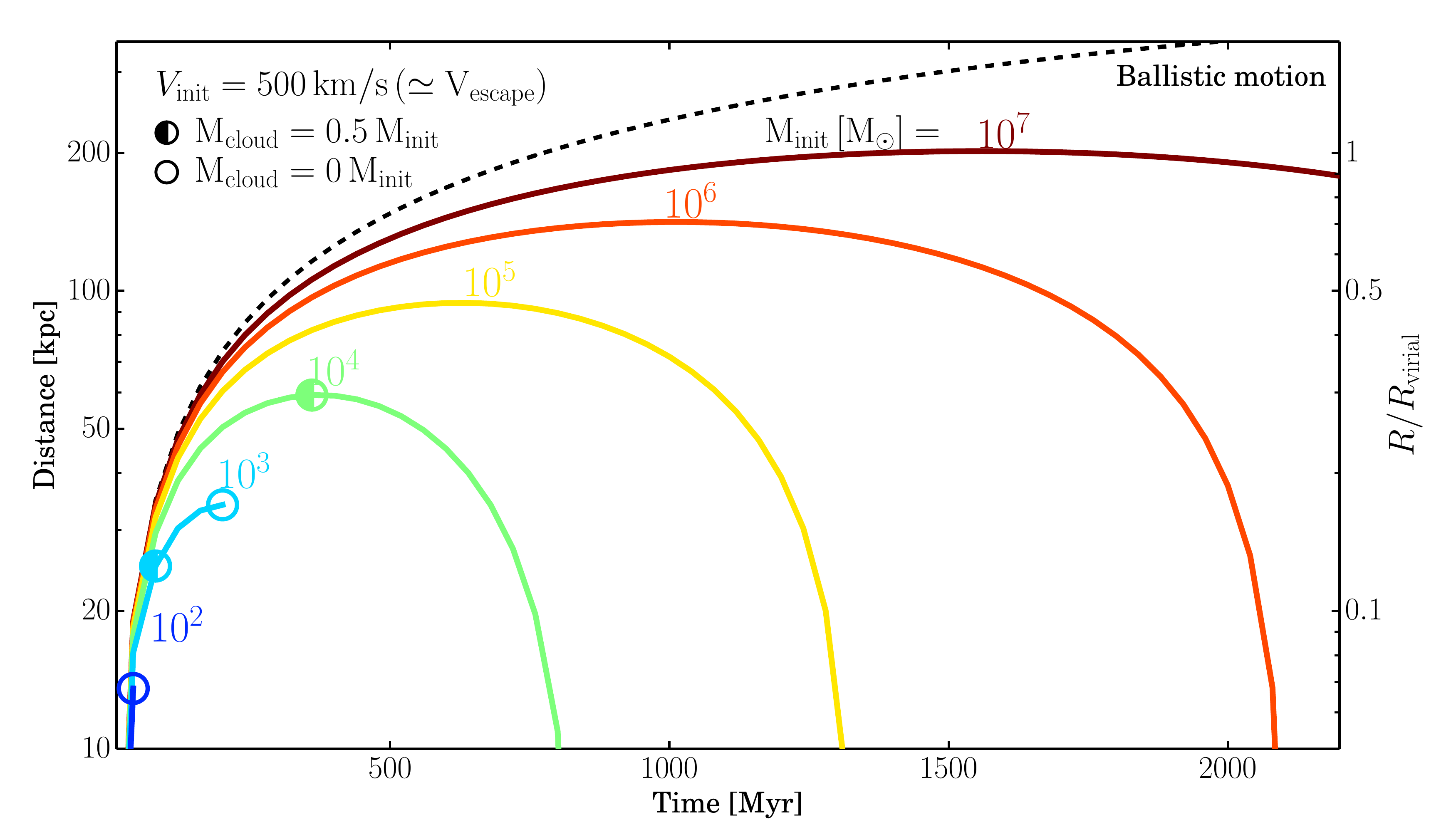}
\caption{Examples of trajectories of clouds with initial velocity of 
$500\,{\rm km/s}$ (escaping velocity) and initial mass from $10^{2}$ to 
$\rm 10^{7} \, M_{\odot}$. Half fill cycles (empty cycles) indicate 
the times when clouds loss half (all) of their initial mass due to 
evaporation. The dashed line shows the trajectory without the effect of 
hot gas. The hot gas prevents the escape of clouds with ram pressure and 
destroys the small cool clouds with evaporation.}
\label{fig:illustration_trajectories}
\end{figure*}

\subsection{Model components for an ensemble of clouds}

To investigate how the statistical properties of 
cool clouds, such as their spatial distribution, depend on the origin of 
the clouds, we populate an ensemble of clouds in a halo by considering the 
following model components:
\begin{itemize}
\item \textbf{Origins of cool gas clouds -} 
    we consider cool gas clouds have three 
    origins: (1) outflow gas clouds ejected at 1 kpc from the center of the halo, 
    (2) inflow gas from the virial radius of the halo, and (3) in-situ gas formed in the 
    halo with distribution following NFW and power law profiles. 
\item \textbf{Gas flow mass rate -} 
    we consider a burst gas flow and a constant
    gas flow for the outflow models.  Mass outflow rate for the burst model follows an 
    exponential form similar to the star-formation history of galaxies 
    \citep[e.g.][]{BC2003}:
    \begin{equation}
        \frac{dM_{\rm outflow}}{dt} =  340 \ e^{-t/200 \rm Myr} \ \rm M_{\odot}/yr.
    \end{equation}
    For the constant outflow and inflow models, we assume a constant gas flow with 20 
    $\rm M_{\odot} \,yr^{-1}$. 
    These models eject $\rm 8\times10^{10} \ M_{\odot}$ in total over 
    4000 Myr, and the value is chosen to produce the cool CGM mass of 
    about $10^{10} \, \rm M_{\odot}$, similar to that observed. We note that there 
    is a degeneracy between the gas flow rate and the evaporation suppression 
    factor $f$ for the evaporation, as discussed in Section 4.  
\item \textbf{Initial cloud mass function -} 
    to populate the total gas mass into 
    clouds, we introduce an initial cloud mass function to account for possibly 
    coverage of cloud masses. We assume that (i) the mass spectrum of cool gas 
    clouds follows a power law distribution with a power $\alpha$:
    \begin{equation}
        N(M_{\rm cloud}) \propto M_{\rm cloud}^{\alpha},
    \end{equation} 
    (ii) the mass spectrum follows a Schechter function with a power $\alpha$ and a 
    characteristic cloud mass $M_{*}$:
    \begin{equation}
        N(M_{\rm cloud}) \propto M_{\rm cloud}^{\alpha}\times e^{-M_{\rm cloud}/M_{*}}.
    \end{equation}     
    In addition, we put limits on the initial maximum and minimum cloud masses. The 
    maximum cloud mass is $10^{8} \ \rm M_{\odot}$ above which the clouds 
    collapse due to self-gravity. The minimum cloud is set to be $10^{2} \ \rm 
    M_{\odot}$. The initial cloud mass function plays an 
    important role in the structure of the 
    circumgalactic medium. The value of $\alpha$ determines how the total cloud mass
    is distributed in cloud mass and, therefore, the column density 
    $N_{\rm HI}$ distribution in the CGM and the covering fraction of gas 
    clouds around galaxies. 
    We will pay special attention to how the properties of the CGM depend on 
    $\alpha$. 
\item \textbf{Velocity distribution - } 
    we assume that the velocity distribution of outflow clouds 
    follows a normal distribution with mean velocity equal to
    $300$, $500$, or $700\,{\rm  km/s}$, and a width of 
    $200\,{\rm  km/s}$. These numbers are motivated by observations 
    \citep[e.g.][]{Rubin2014} and theoretical models \citep[e.g.][]{Murray2011}. 
    Clouds are assumed to be ejected from the center isotropically.
\end{itemize}

%%%%%%%%%%%% Section 3 %%%%%%%%%%%%%%%%%%%%%%%%%%%%%%%%%

\section{Results}

%======================================================
%Figure 
%======================================================
\begin{figure*}[t]
\center
\includegraphics[scale=0.27]{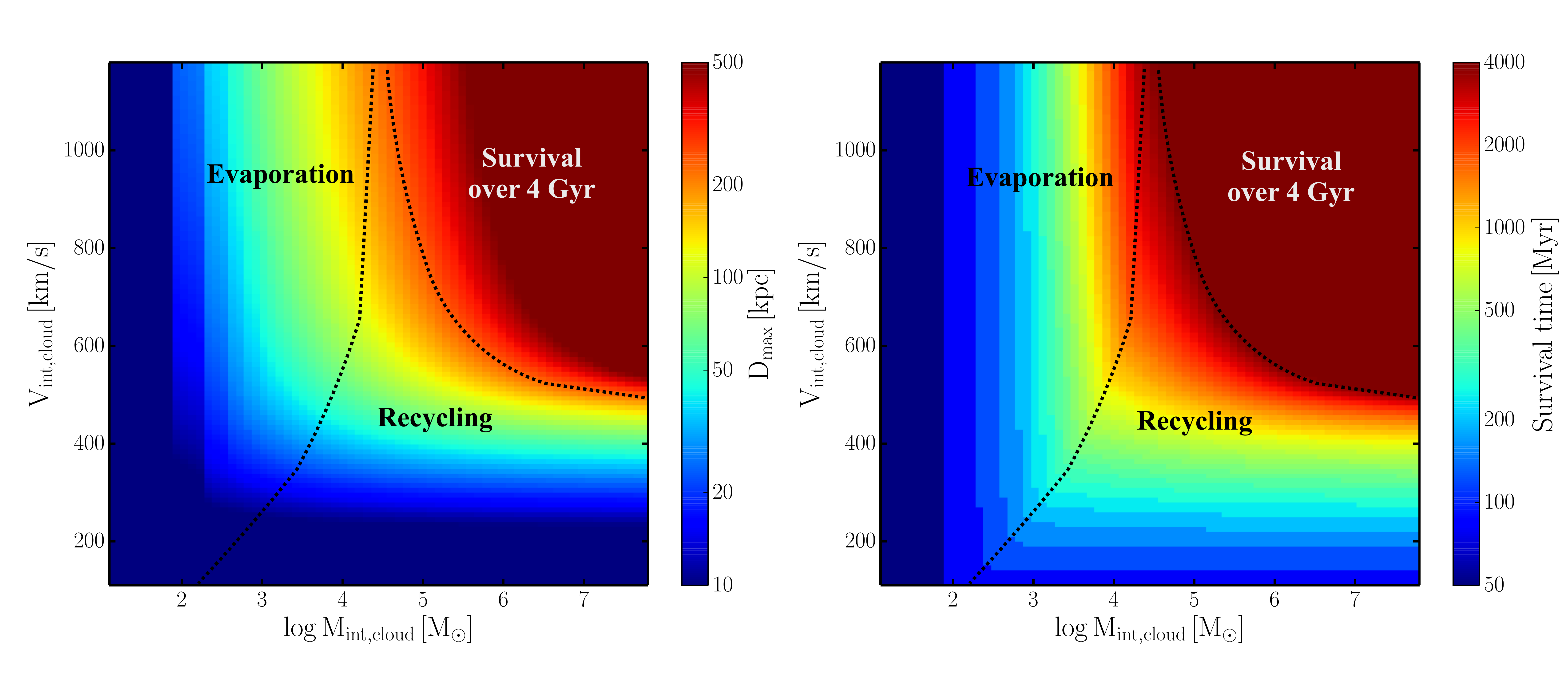}
\caption{Summary of the fates of outflowing clouds with various 
initial masses and velocities. 
\emph{Left:} maximum distance. \emph{Right}: survival time.}
\label{fig:summarize_final_state}
\end{figure*}
%======================================================

%======================================================
%Figure 
%======================================================
\begin{figure}[t]
\center
\includegraphics[scale=0.335]{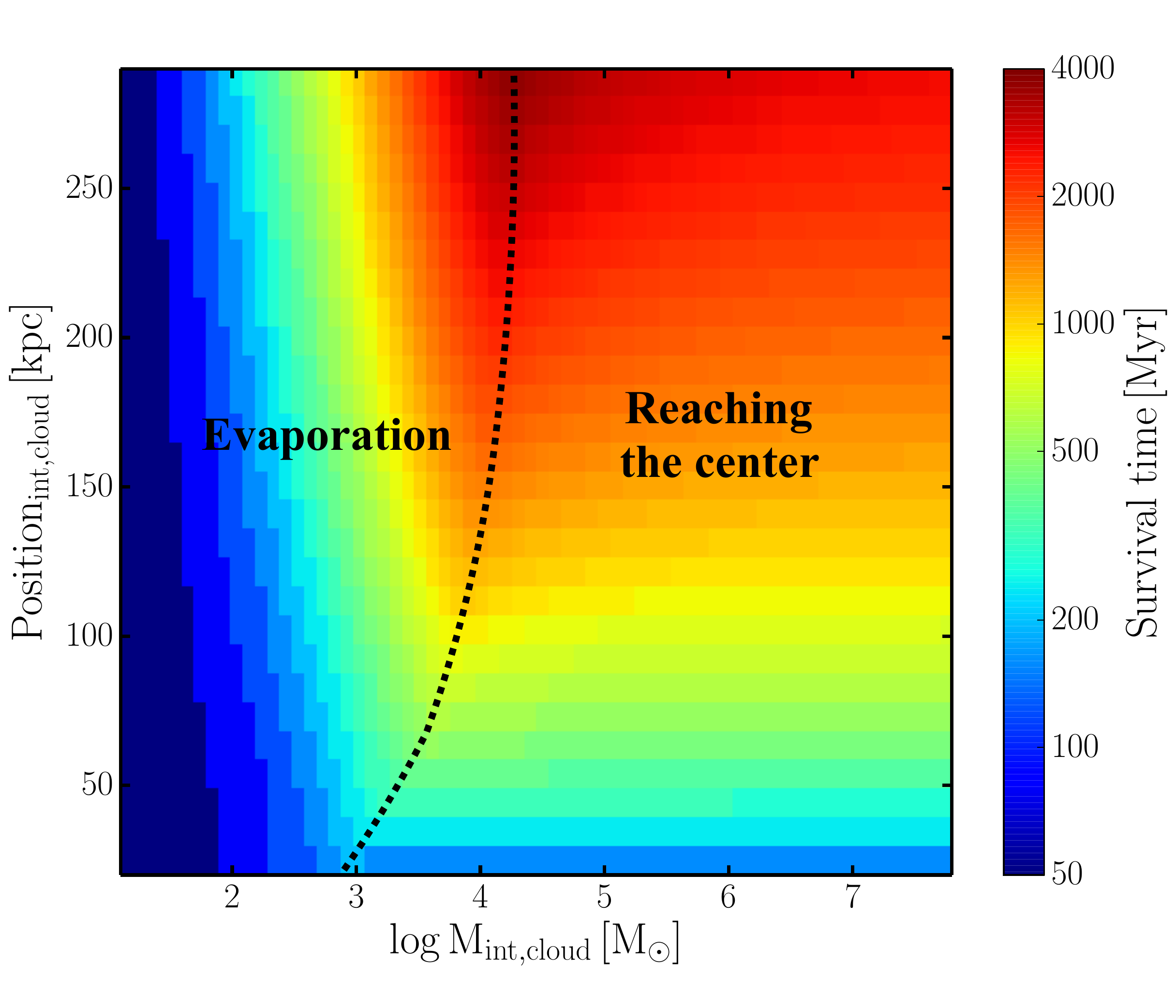}
\caption{Summary of the fates and survival time of inflow clouds with various initial masses and positions.}
\label{fig:summarize_final_state}
\end{figure}
%=

%
%
%======================================================
%Figure 
%======================================================
\begin{figure*}[t]
\center
\includegraphics[scale=0.4]{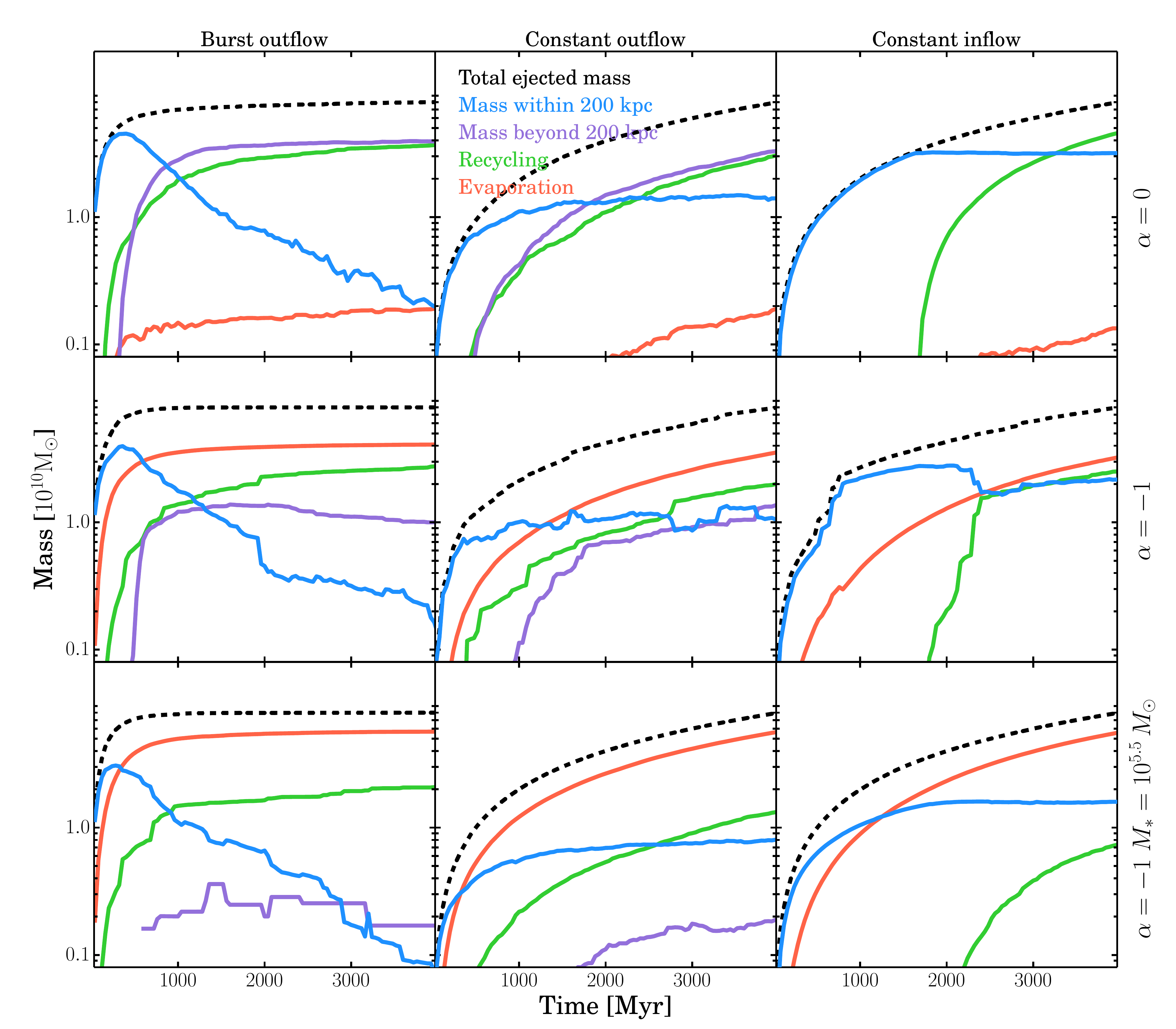}
\caption{Mass evolution of ejected clouds with three gas flow modes. 
\emph{Left:} Burst outflow models; 
\emph{Middle}: constant outflow models; 
\emph{Right}: constant inflow models. 
The total ejected mass is shown with the black lines. 
The masses in the circumgalactic space and moving beyond the virial radius 
are shown in blue and purple, respectively.
The mass that recycles back to the center is shown in green and 
that gets evaporated is shown in red.}
\label{fig:mass_budget}
\end{figure*}

\subsection{Evolution of individual clouds}

We first study the motion and mass evolution of clouds derived from 
Equations~(\ref{eq:motion}) and (\ref{eq:mass_loss_saturation}).
Figure 3 shows examples of the trajectories of clouds with masses 
from $\rm 10^{2}$ (blue line) to $10^{7} \, \rm M_{\odot}$ (red line), 
and with initial velocity $V_{\rm init}=500$ km/s ejected at 1 kpc from 
the center of the halo. Ballistic motion without the effect of ram pressure 
is shown with the black dashed line. The half-filled and open circles 
indicate the times when the cloud has lost half of its initial 
mass and is totally evaporated, respectively. As shown in the figure, the 
effect of ram pressure depends on the mass of clouds: massive clouds 
are able to travel beyond 100 kpc while small clouds are stopped by ram 
pressure and fall back within 50 kpc. Without ram pressure, clouds with 
the same initial velocity should all follow the ballistic motion. 
For cloud masses $<10^{7} M_{\odot}$, no clouds can escape the halo 
even if the initial velocity is comparable to the escape velocity of 
the gravitational potential. In addition to ram pressure, the hot 
ambient gas can also remove gas from clouds via heat evaporation. 
Clouds with masses less than $10^{4} \ \rm M_{\odot}$ lose all the 
mass within a few hundred Myr due to evaporation, while clouds with 
masses greater than $10^{4} \ \rm M_{\odot}$ fall back to the center 
with a fraction of their initial masses. The estimated evaporation 
timescales for clouds are similar to the timescale estimation 
from hydrodynamical simulations by \citet{Armillotta2017} for individual clouds.

We summarize the fates of clouds with various initial velocities and masses 
ejected at 1 kpc from the halo center in Figure 4. The left panel shows the maximum distance 
that a cloud can travel and the right panel shows the 
survival time of the clouds. The fates of clouds can be classified 
into three categories: 
\begin{enumerate}
\item Evaporation (left): 
    Clouds with relative low masses and high 
    velocities lose all their masses due to evaporation within 50-1000 Myr 
    while traveling in the halo. Even with extreme initial velocities, small 
    clouds can only reach about 50-100 kpc from the center before being evaporated.  
\item Survival (top right): 
    Clouds with high masses and high velocities can 
    survive over 4 Gyr and most of them can escape the halo. 
\item Recycling (bottom right): 
    Clouds with high masses but low velocities 
    eventually fall back to the halo center with a fraction of their initial 
    masses. The timescale for clouds to be recycled is about 500-1000 Myr 
    depending on the initial velocity consistent with results from hydrodynamical simulations \citep[e.g.,][]{Ford2014,Afire2017}.
\end{enumerate}

In addition to clouds ejected from the halo center, we also model the gas clouds falling from a given initial position towards the halo center with no initial velocity. 
The fates of the inflowing gas clouds are summarized in Figure 5 showing the survival time of the clouds. Similarly to the 
outflow case, clouds with low masses are evaporated in the circumgalactic 
space while massive clouds can reach the halo center with a fraction of 
their initial masses. For clouds falling from the halo virial radius 
($\sim$200 kpc), it takes about 2 Gyr to reach the center.

\subsection{Mass Evolution}

With the properties and fates of individual clouds modeled, we now explore how the ensemble of clouds evolves with time. 
To do so, we eject clouds with a given flow rate and a given 
initial cloud mass function, as discussed in Section 2.2. 
We estimate the fractions of initial ejected mass (1) staying in the 
circumgalactic space, (2) moving beyond the virial radius, (3) evaporated, 
and (4) falling back into the center, and explore how the different 
fractions change with model parameters. Figure 6 shows the results for 
three modes of gas flow models: the burst outflow model (left column), 
constant outflow (middle column), and constant inflow models (right column). 
We also explore three initial mass functions, power laws with 
$\alpha=0$ (top panel) and $\alpha=-1$ (middle panel), and a 
Schechter function with $\alpha=-1$ and $M_{*}=10^{5.5} \, M_{\odot}$ 
(bottom panel). The color lines indicate masses in different components. 
Mass within and outside the virial radius are shown with the blue and purple 
lines, respectively. The green line indicates the recycled mass that falls 
back to the center, and the red line shows the amount of mass that is 
evaporated. The black dashed line shows the total ejected mass. The gas flow rate 
is selected to eject $8\times10^{10} M_{\odot}$ over 4 Gyr. The mean initial 
velocity for the outflow models is 500 km/s.

Let us first focus on the burst outflow model with $\alpha=0$, shown in  
top-left panel, in which all the mass ($8\times10^{10}\rm \, M_{\odot}$) 
is ejected in 500 Myr. The mass in the circumgalactic medium reaches its 
maximum at around 400 Myr, and declines afterwards. After 1 Gyr, 
$\sim 40\%$ of the total mass has moved beyond the virial radius and a 
similar fraction of the mass recycled back to the center. Only a small 
fraction of mass remains in the halo ($15\%$) or gets evaporated ($5\%$). 
This demonstrates that a single burst event cannot sustain the circumgalactic 
medium for a timescale of billion years. In contrast, in the constant outflow 
model, shown in the top middle panel, the mass in the circumgalactic region 
can reach $\sim 10^{10} \rm M_{\odot}$ and remains more or less constant.   
This is due to the balance between the constant mass outflow, the 
recycled mass, and the escaped mass. As in the burst outflow, $\sim40\%$ of the ejected 
mass returns to the center as recycled mass, while $\sim40\%$ escapes the halo. 
The mass evolution of the constant inflow model behaves similarly to the constant 
outflow model except that it takes about 2000 Myr for the inflowing clouds 
to reach the halo center. 

This mass evolution also depends on the shape of the initial mass function:
the typical cloud mass decreases and the evaporation timescale becomes 
shorter as $\alpha$ decreases. By changing the shape of the initial mass 
function from $\alpha=0$ to $\alpha=-1$, the amount of evaporated mass increases
by a factor of 10, as shown in the middle panels of Figure 6. Finally, adopting a 
Schechter function with $\alpha=-1$ and $M_{*}=10^{5.5} \, M_{\odot}$ as the 
initial cloud mass function further reduces the number of massive clouds and 
enhances the mass fraction of evaporation to about $80\%$, as shown in the 
bottom panels of Figure 6. One interesting implication is that heat evaporation 
reduces the total cloud mass from outflows. Therefore, the outflow mass estimated from the cool CGM mass is only a lower limit \citep[e.g.,][]{Lan2018} and it can be an order of magnitude lower than the intrinsic outflow mass depending on the shape of the mass function.  
Similarly, the evaporation reduces the total infalling cloud mass that reaches the center. This 
might be a possible mechanism to prevent further star-formation in 
passive galaxies \citep[see also][]{Zahedy2018}.
These results demonstrate that the total mass in the 
CGM depends on the shape of the initial cloud mass function. In the following, we will 
show that the shape of the initial mass function also plays an important  
role in determining the profile of the circumgalactic medium. 
%======================================================
%Figure 
%======================================================
\begin{figure*}[t]
\center
\includegraphics[scale=0.45]{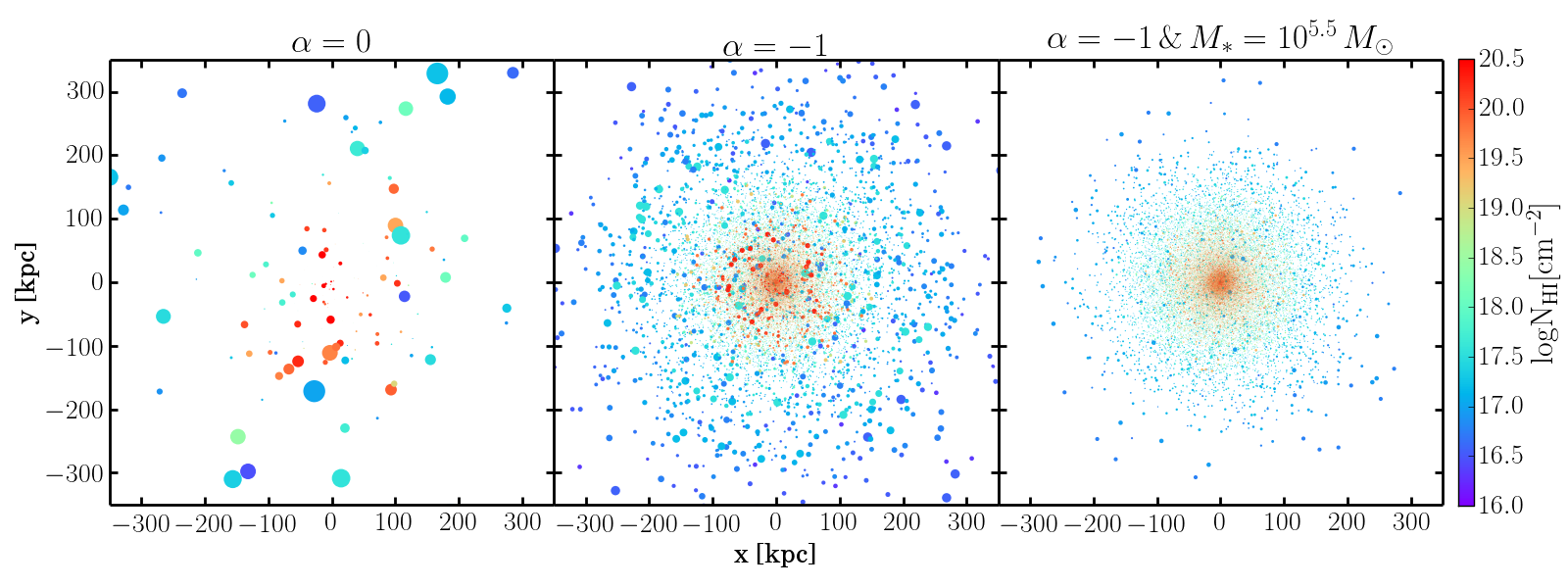}
\caption{Snapshots of the spatial distribution of gas clouds in the 
constant outflow models with three different initial cloud mass 
functions at 2 Gyr. \emph{Left:} a power law initial mass function with 
$\alpha=0$. \emph{Middle}: a power law initial mass function with 
$\alpha=-1$. \emph{Right:} a Schecther-like initial mass function with 
$\alpha=-1$ and $M_{*}=10^{5.5} \, \rm M_{\odot}$. 
Only $10\%$ of the clouds are plotted with sizes reflecting 3 times 
of the physical sizes of the clouds. The color indicates its $\rm N_{HI}$. 
The mean initial velocity is 500 km/s.}
\label{fig:cloud_distribution}
\end{figure*}
%======================================================

\subsection{Spatial distribution of cool gas clouds}

We now investigate the spatial distribution of the pressure-confined gas clouds as a 
function of different initial mass function and initial velocity. To guide the exploration, 
we compare our modeled gas properties with the observed neutral hydrogen gas 
around star-forming and passive galaxies at redshift 0.5. To this end, we use the 
distribution of the neutral hydrogen traced by strong MgII absorbers. 
Such absorbers, with $W_{2796}>0.4\ \rm \AA$, have been considered as a tracer 
of the cool circumgalactic gas with neutral hydrogen column densities
$N_{\rm HI}>10^{18.5} \rm \ cm^{-2}$ and volume densities $n_{\rm H}\sim0.3 \ \rm cm^{-3}$ 
\citep[e.g.,][]{Rao2006,Menard2009,Lan2017}, consistent with the properties of 
the pressure-confined cool clouds in our model. 
Strong MgII absorbers have been observed around both star-forming and 
passive galaxies, and their distributions around both types of 
galaxies are similar on large scales ($r_{p}>50$ kpc) 
\citep[e.g.,][]{Chen2010,Nielsen2013,Lan2014,Lan2018}. 
Within 50 kpc, however, there is more MgII absorption around star-forming 
galaxies than around passive galaxies. This excess absorption is most likely 
associated with outflows, as is inferred from the azimuthal angle 
distribution \citep[e.g.,][]{Bordoloi2011,Kacprzak2012,Lan2018}. 
In our analysis, we use the MgII covering fraction at redshift 0.5 around galaxies with stellar mass $\rm \sim 10^{10.8} \, M_{\odot}$
provided by \citet{Lan2014} with 
$0.4<W_{\lambda2796}<0.8 \,\rm \AA$,  $0.8<W_{\lambda2796}<1.5\, \rm \AA$, 
and $W_{\lambda2796}>1.5 \, \rm \AA$, as the proxy of neutral 
hydrogen covering fraction with $\rm 18.6<log \, N_{HI}/cm^{-2}<19.1 $,  
$\rm 19.1<log \, N_{HI}/cm^{-2}<19.6 $, and $\rm log \, N_{HI}/cm^{-2}>19.6$, 
respectively,  based on Equation (2).

Our goal is to explore how different physical mechanisms affect 
the gas properties by using observational measurements as a guidance. 
To this end, we will examine the effects of individual model parameters one by one.

\subsubsection{Constant outflow models}

In this subsection, we explore how the ensemble properties of cool gas clouds 
depend on the initial cloud mass function and the initial cloud velocity by using 
constant outflow models as examples.

\textbf{The effects of initial cloud mass function} are illustrated in Figure 7, 
which shows the structure of the circumgalactic medium as a function of the shape of 
the initial cloud mass function. Here the mean initial cloud velocity is set to be 
$500$ km/s and a constant outflow rate of 20 $\rm M_{\odot}/yr$ is assumed. 
Results are shown for the snapshot at 2 Gyr. We show only $10\%$ of total number
of clouds with sizes of data points reflecting 3 times of the physical sizes of clouds.  
The total mass in the cool circumgalactic medium is about $10^{9.5}-10^{10} \,M_{\odot}$, 
with the exact value depending on the shape of the initial mass function. 
The color indicates the $\rm N_{HI}$ the gas cloud.  

The morphology and structure of the cool CGM both depend on the initial cloud 
mass function. For a power law function with $\alpha=0$, the CGM consists of 
thousands of massive clouds with $\rm N_{HI}>10^{20} \, cm^{-2}$ (left panel), 
while the CGM produced by a power law mass function with $\alpha=-1$ 
contains millions of gas clouds with $\rm N_{HI}$ ranging from $10^{18.5}$ to 
$10^{20.5}\rm \,cm^{-2}$ (middle panel). With an exponential cutoff in the 
massive end, the Schechter initial mass function further reduces the number of 
massive clouds (red dots) and produces even more small clouds (right panel). 

\begin{figure*}[t]
\center
\includegraphics[scale=0.5]{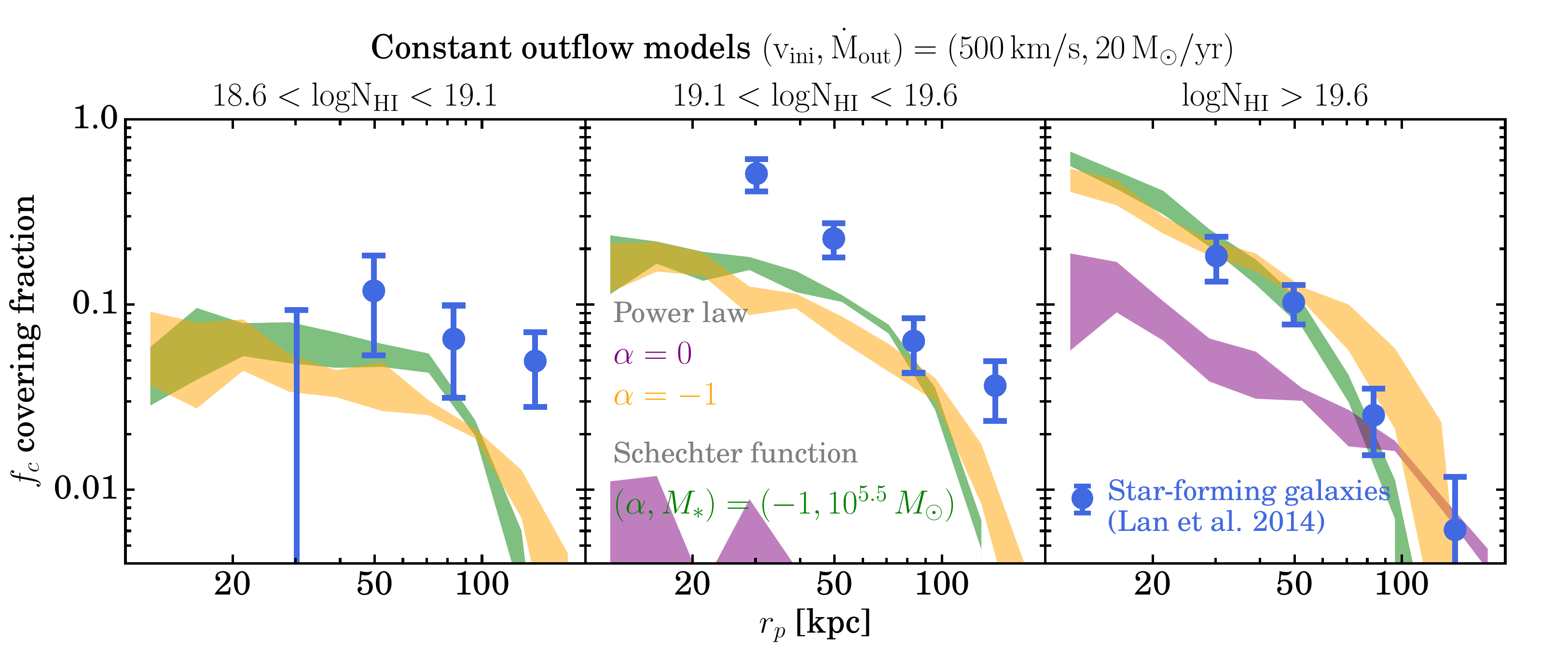}
\includegraphics[scale=0.5]{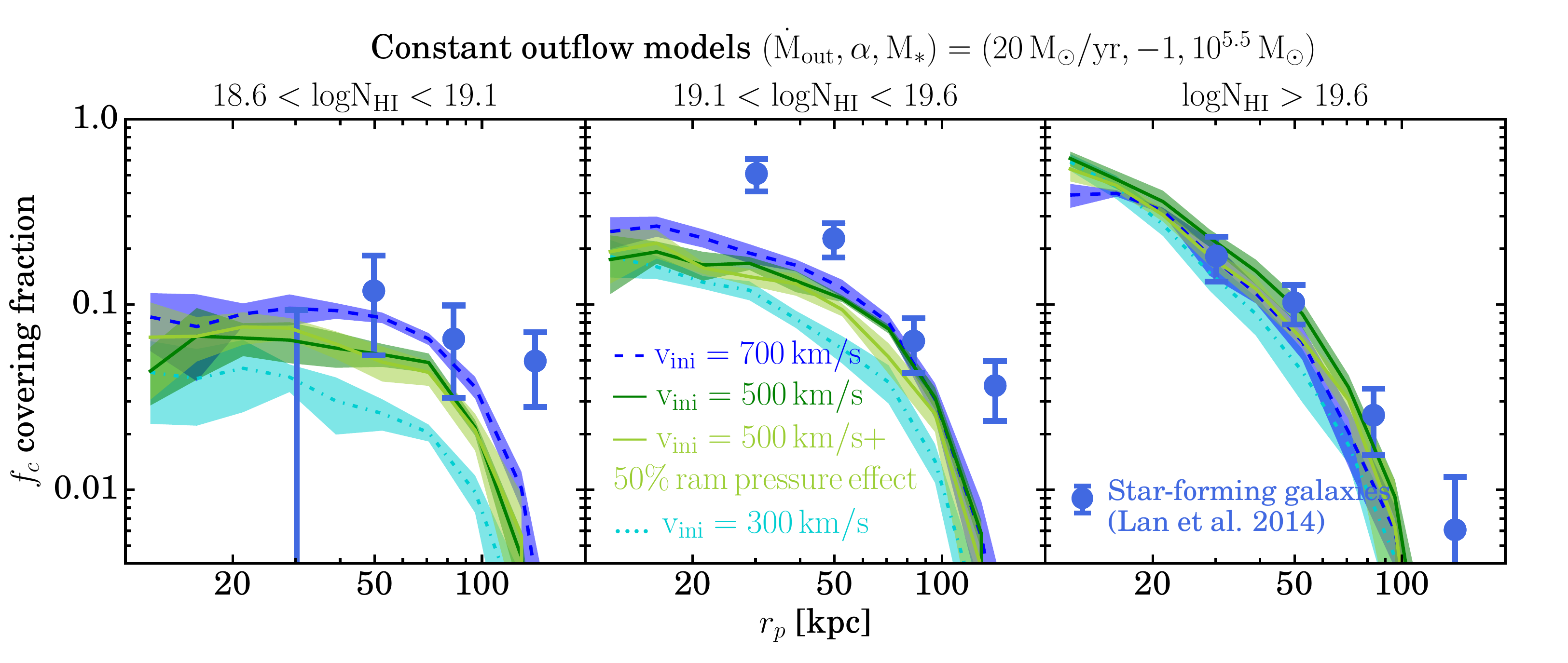}
\includegraphics[scale=0.5]{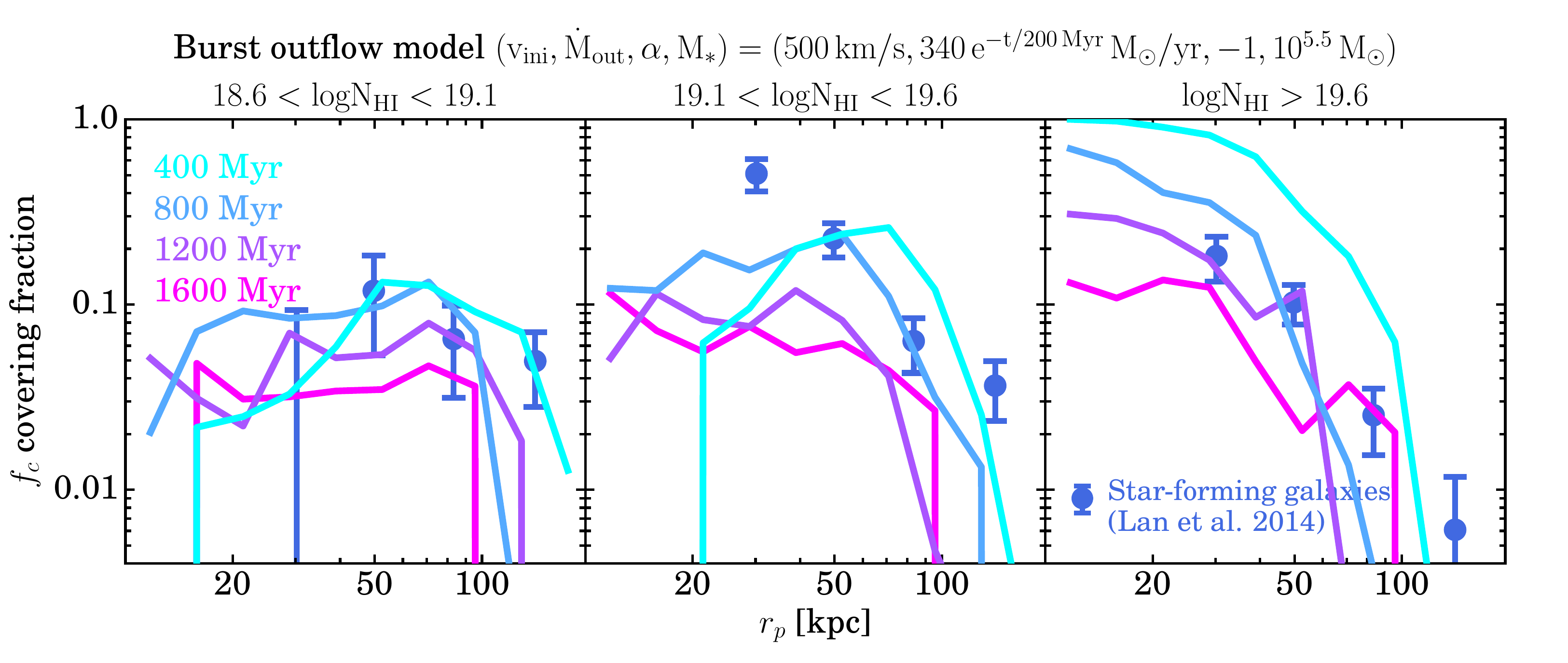}

\caption{Covering fraction of gas clouds as a function of neutral hydrogen column 
density. The blue data points show the observed covering fraction from Lan et al. (2014).
\emph{Top:} Covering fraction as a function of initial cloud mass function. The results of 
power law mass functions with $\alpha=0$ and $\alpha=-1$ are shown by purple and 
orange lines respectively. The green lines show the results from a Schechter 
like mass function with $\alpha=-1$ and $M_{*}=10^{5.5} \rm \, M_{\odot}$. 
\emph{Middle:} Covering fraction as a function of mean initial cloud velocity 
with 300, 500, and 700 km/s shown by dotted, solid, and dashed lines, respectively. 
\emph{Bottom:} Covering fraction as a function of time with a burst outflow model. 
}
\label{fig:outflow_cf}
\end{figure*}

To quantify the effects of the initial cloud mass function, we estimate the covering 
fraction of gas clouds as a function of $\rm N_{HI}$ and compare the results 
with the observed ones in the top panel of Figure~\ref{fig:outflow_cf}. 
The predicted covering fractions are estimated from the interceptions 
with clouds out of 50,000 random sightlines with impact parameters $\le 250\,{\rm kpc}$. 
When multiple clouds are intercepted by a single sightline, the $\rm N_{HI}$ 
is the sum of the $\rm N_{HI}$ of individual intercepted clouds.
The average neutral hydrogen gas covering fraction over the time interval 
between 2000 Myr and 4000 Myr produced by the power law mass functions with 
$\alpha=0$ and $\alpha=-1$ are shown in purple and orange, respectively, 
with the bands showing the fluctuation of the covering fraction over time. 
With $\alpha=0$, the gas covering fraction of $\rm N_{\rm HI}<10^{19.6} \, \rm cm^{-2}$
clouds is close to zero as almost all the outflow mass is carried by massive clouds, 
inconsistent with the observed covering fraction shown by the blue data points. 
On the other hand, the number of massive clouds is reduced and that of small clouds 
enhanced in the $\alpha=-1$ model, and the predicted covering fraction of 
$\rm N_{\rm HI}<10^{19.6} \, \rm cm^{-2}$ is now of the same order as 
the observed one.  
%
%======================================================
%Figure 
%======================================================
%======================================================
%

%
%
%======================================================
%Figure 
%======================================================
\begin{figure*}[t]
\center
\includegraphics[scale=0.43]{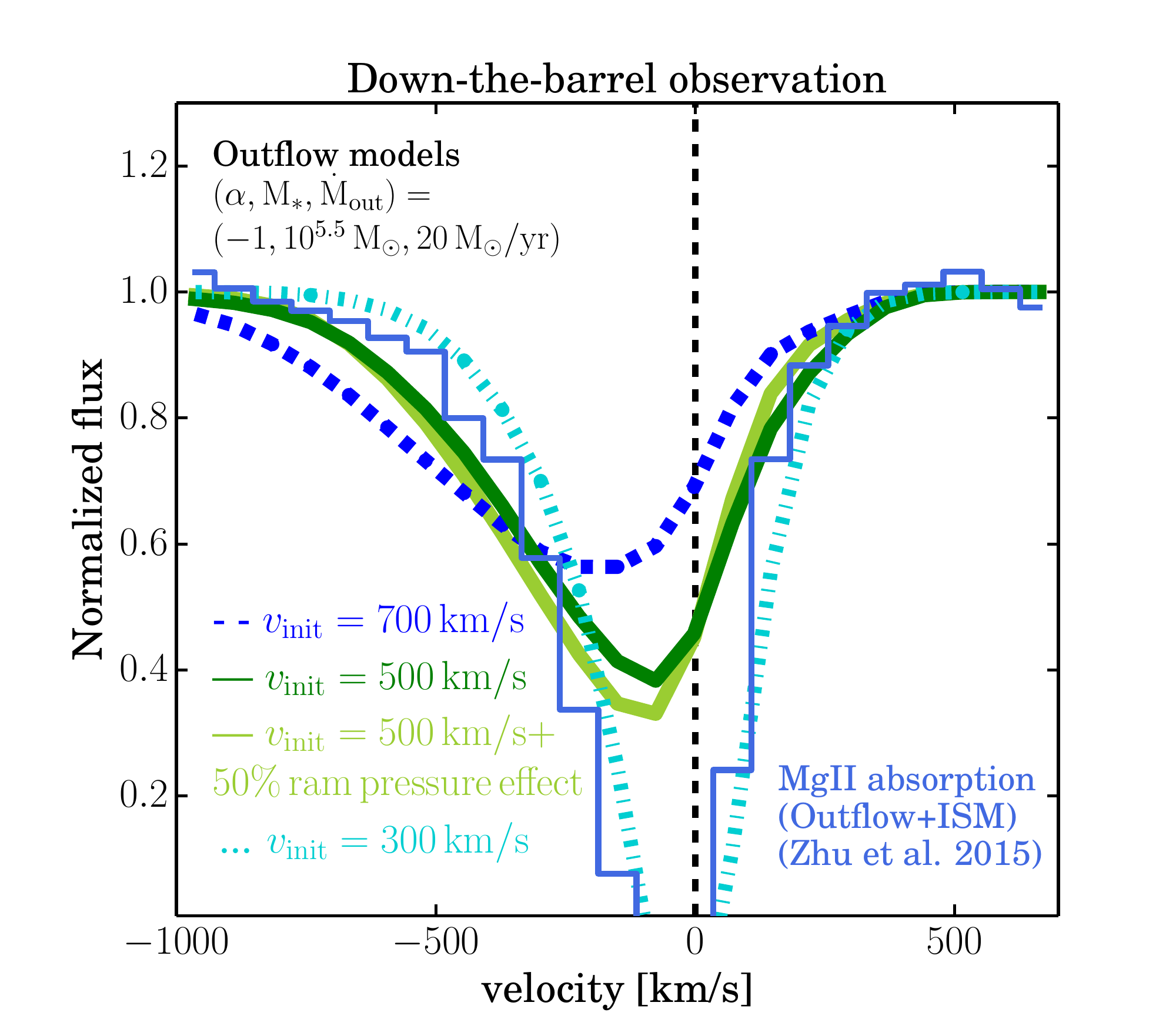}
\includegraphics[scale=0.43]{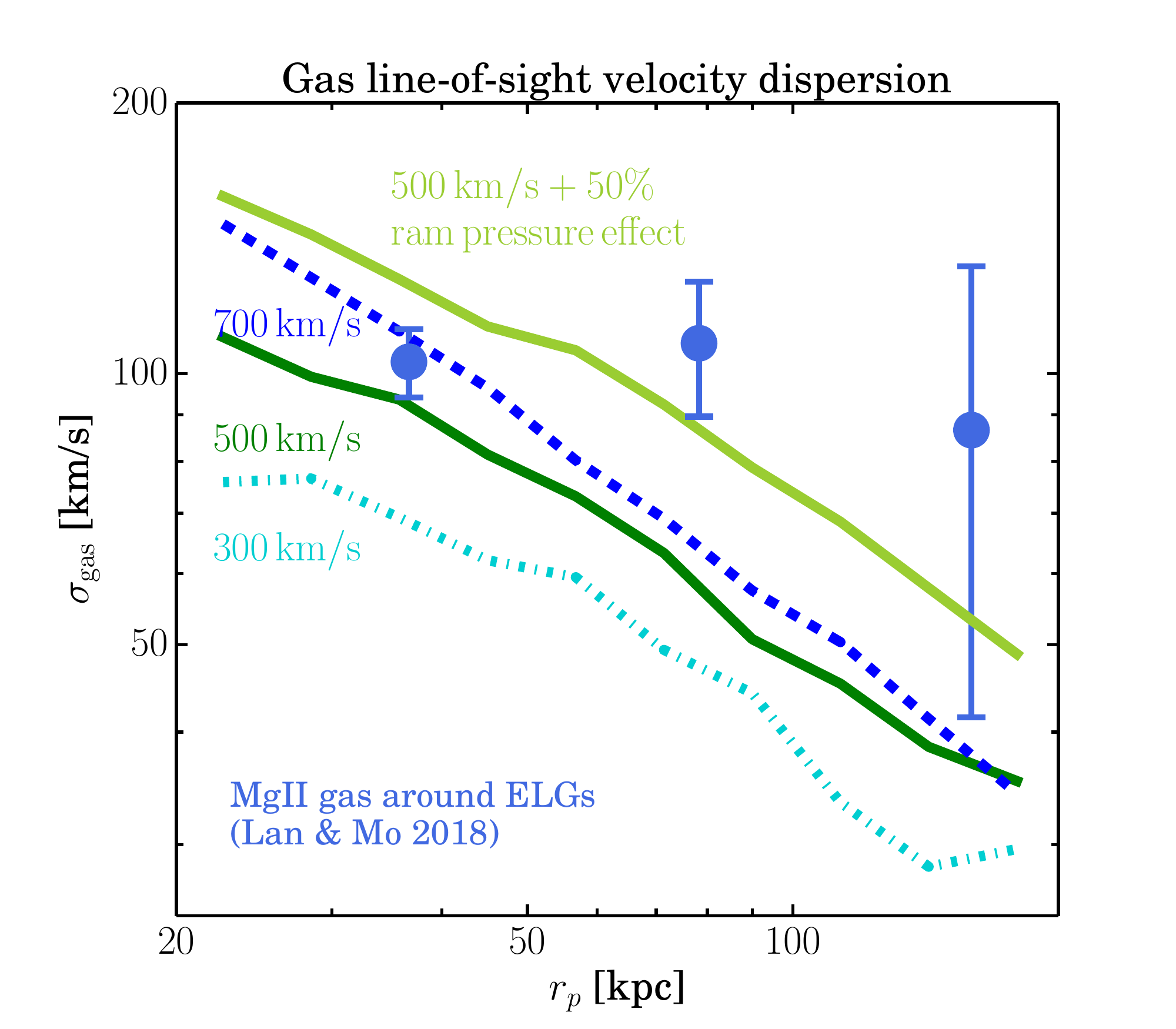}
\caption{\emph{Left:} Down-the-barrel blue-shifted outflow absorption. 
The blue histogram shows the observed down-the-barrel MgII absorption from 
\citet{Zhu2015} with emission-line filling corrected.  \emph{Right:} The line-of-sight velocity 
dispersion. The observed line-of-sight velocity dispersion of MgII gas from 
\citet{Lan2018} is shown by the blue data points. In both panels, the lines 
show the modeled properties with initial velocity 300 (cyan dotted), 
500 (green solid), and 700 (blue dashed) km/s. The light green line shows the model with initial velocity 500 km/s and $50\%$ of the ram pressure efficiency.}
\label{fig:velocity}
\end{figure*}
%======================================================
%

Finally, the results assuming a Schechter function with $\alpha=-1$ and 
$M_{*}=10^{5.5} \, \rm M_{\odot}$ for the cloud mass function are shown in green
color. The Schechter function redistributes the mass in massive clouds 
($M_{\rm cloud}>M_{*}$) into smaller ones, thereby further enhancing the gas 
covering fraction of systems with $\rm 10^{18.6}<N_{HI}<10^{19.6} \rm cm^{-2}$. 
The difference between the orange (power law) and green lines (Schechter function) 
illustrates the effect of the mass function. Such an effect can also be seen from the 
spatial distribution shown in Figure 7. These results demonstrate that the 
initial cloud function can affect the structure the gas distribution around galaxies
significantly, and so can be constrained by the observed spatial gas distribution as a 
function of column densities. As one can see, the outflow model with the Schechter cloud 
mass function can reproduce roughly the observed gas covering fraction as a 
function of neutral hydrogen column densities.

\textbf{The effects of initial outflow velocity} are shown in the middle row of 
Figure~\ref{fig:outflow_cf}, where results obtained from  three initial outflow mean 
velocities, 300 (cyan dotted), 500 (green solid) and 700 (blue dashed lines) km/s
are plotted. 
Comparing to the model with 500 km/s, the model with 300 km/s 
is insufficient for the clouds to propagate far enough. On the other hand, 
increasing the mean velocity to be 700 km/s will allow more clouds to   
propagate to larger distances. However, since the value of $N_{\rm HI}$
for a fixed cloud mass decreases rapidly at large distance, as shown in Fig. 2, 
the clouds that propagate to large distances do not contribute significantly 
to the covering fraction of the relatively high-$N_{\rm HI}$ systems concerned 
here. 

The initial outflow velocity in a galaxy may be constrained by the gas absorption 
observed toward the galaxy, i.e. through the so-called down-the-barrel observation. 
To do this, we assume that the effective radius of a galaxy is 5 kpc and calculate the 
absorption profile based on 
\begin{equation}
    \frac{I}{I_{0}}(v)=1-f_{c}(v)\times(1-e^{-\tau}),
\end{equation}
where I and $I_{0}$ are the observed flux and the continuum, respectively, 
$f_{c}(v)$ is the covering fraction of clouds at a given velocity, and $\tau$ is 
the optical depth \citep[e.g.][]{Jone18}. Here we consider the optical thick region 
where absorption lines are saturated and $\tau$ is large. In this case
the above equation can be approximated as $1-f_{c}(v)$. The left panel of Figure 9 
shows our modeled down-the-barrel absorption. The cyan dotted, green solid, and blue dashed lines show the 
results with initial outflow velocity equal to 300, 500, and 700 km/s, respectively. 
For comparison, we also show the observed down-the-barrel strong MgII 
absorption of star-forming galaxies (blue histogram)
from \citet{Zhu2015}. 
We note that the observed profile includes not only the absorption from outflowing gas but 
also from interstellar gas,  while the modeled profile includes only outflowing gas. 
As the interstellar gas absorption is expected to occur mostly around zero velocity, 
the observed blue-shifted absorption wing ($<-200$ km/s) is expected to be due to 
outflowing gas, and so can be directly compared with the modeled profile. 
The models assuming a mean initial velocity between 300 and 500 km/s 
is able to roughly reproduce the blue-shifted absorption profiles seen in the 
observation, while the model with 700 km/s produces absorption that is stronger 
than the observed one at velocities $<-500$ km/s. The modeled down-the-barrel absorption 
profile is asymmetric, consistent with the observation. This is due to two effects 
of the hot gas: (1) the hot gas evaporates a fraction of outflow clouds as 
they move out,  and (2) the ram pressure decelerates the clouds even when they 
are falling back towards the center.

Another observable that can be used to test the model prediction is the line of sight 
gas velocity dispersion. To derive such quantities with our models, we first estimate the 
gas covering fraction as a function of velocity and obtain mock absorption spectra as a 
function of impact parameters. To compare and mimic the observation from SDSS \citep{Lan2018}, 
we first smooth the mock absorption spectra with a Gaussian kernel with 70 km/s 
(SDSS spectral resolution) and then estimate the gas velocity dispersion by fitting the 
smoothed mock absorption spectra with a Gaussian profile. Finally, we subtract the spectral 
resolution in quadrature as done in \citet{Lan2018}. The right panel of Figure~\ref{fig:velocity} shows 
the modeled line-of-sight velocity dispersion of pressure-confined outflow gas clouds. 
As expected, the gas velocity dispersion increases with the mean initial outflow velocity. 
The modeled gas velocity dispersion decreases with impact parameter due to the fact 
that the velocities of clouds are decelerated by both the gravity and the ram pressure 
of the hot gas. For clouds with $M_{\rm cloud}\sim10^{4}-10^{5} \, M_{\odot}$ that can reach 
large distances ($\sim 100 \, \rm kpc$, as shown in Figure 3), their velocities are 
only tens of km/s. This is in contrast to the observed gas line of sight 
velocity dispersion obtained from MgII absorption lines detected in the stacked background 
quasar spectra around star-forming galaxies, shown by the blue data points \citep{Lan2018}. 
This indicates that some other mechanisms are needed to increase the gas velocity 
dispersion in the halos. One possible way is to reduce the deceleration 
from the ram pressure effect, as discussed below. 

\begin{figure*}[t]
\center

\includegraphics[scale=0.5]{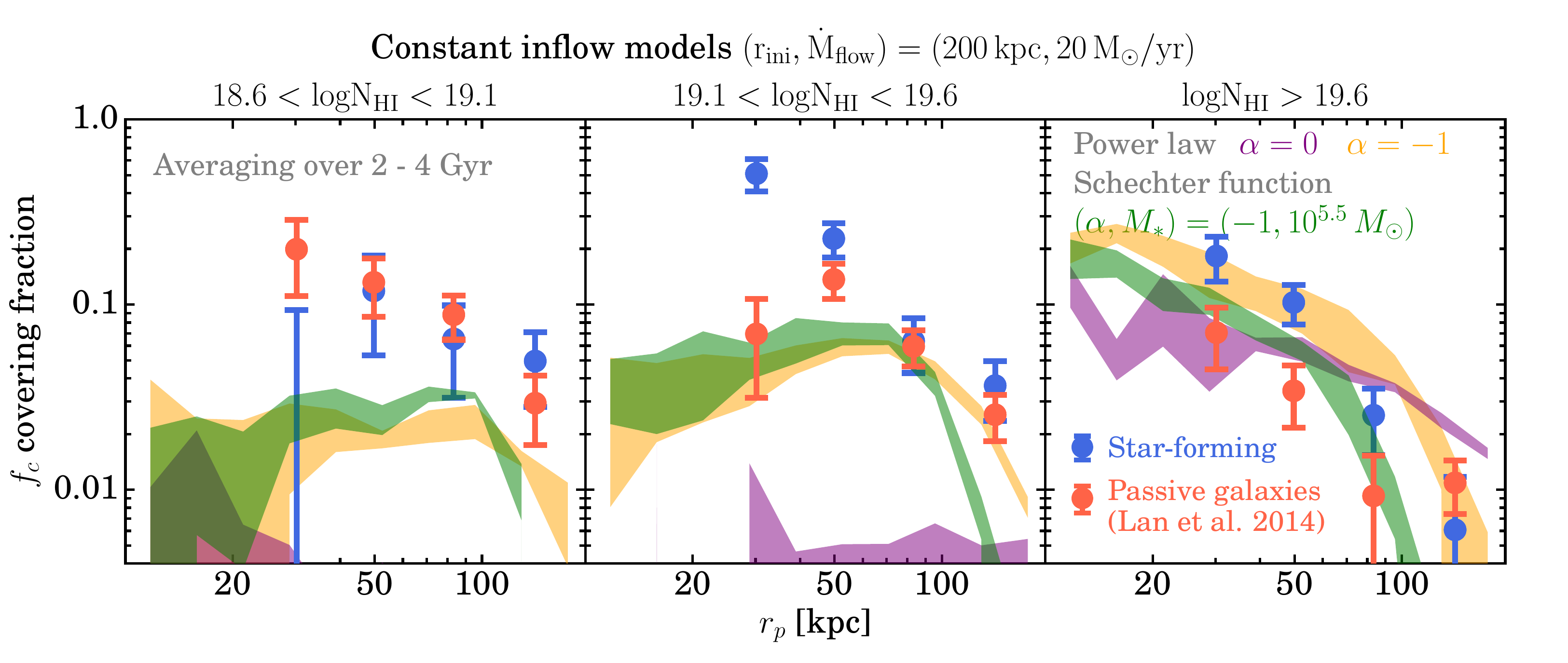}
\includegraphics[scale=0.5]{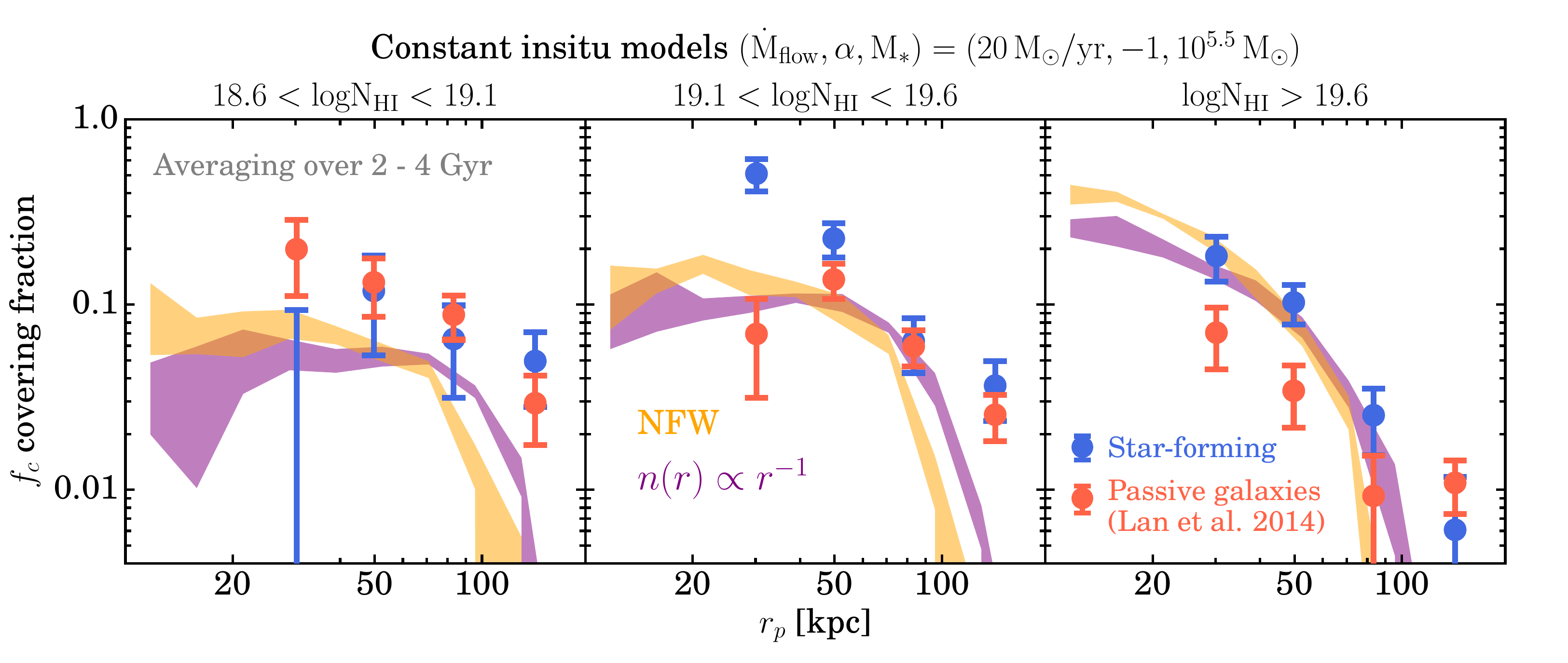}
\caption{\emph{Top:} Predicted covering fractions by inflow models as functions of 
the initial mass function. The results of power law mass functions with $\alpha=0$ and 
$\alpha=-1$ are shown by purple and orange lines, respectively. The green lines show 
the results of a Schechter like mass function with $\alpha=-1$ and 
$M_{*}=10^{5.5} \rm \, M_{\odot}$. 
\emph{Bottom:} Predicted covering fractions by in-situ models with a NFW density profile (orange) and a power law profile (purple).
The data points are covering fractions obtained by \citet{Lan2014}. }
\label{fig:cf_inflow}
\end{figure*}
%======================================================

\textbf{The effects of ram pressure} - To explore how the ram pressure affect the properties 
of the gas, we adjust the ram pressure efficiency $C_{d}$ in Equation 4 to be only $50\%$ of the default value. 
The results are shown by the light green color in Figure 9. As one can see, 
lowering the ram pressure efficiency does not affect significantly the down-the-barrel 
absorption. This is because the absorption is mostly produced by gas clouds 
that have just been ejected from the center (left panel of Figure 9). On the other hand, 
by lowering the ram pressure efficiency, the gas line-of-sight velocity dispersion is enhanced 
by about a factor of 1.5, closer to the observed values, as shown in the right panel of 
Figure 9. 
Lowering the ram pressure efficiency allows more gas clouds to travel to larger 
distances, but it does not affect the gas covering fraction significantly, 
as shown in the middle panel of Figure 8, where the gas 
covering fractions obtained from the default ram pressure efficiency and from $50\%$ 
efficiency are shown by the green and light green lines, respectively.

These results 
indicate that only with a certain combination of parameters, the semi-analytic model 
considered here is able to reproduce the observed spatial distribution, down-the-barrel 
profile, and gas kinematics of cool CGM around star-forming galaxies. 
In other words, these observed gas properties together can provide strong constraints 
on physical models of outflows.

\subsubsection{Burst outflow models}
 In addition to the constant outflow models, 
we show the covering fraction of gas predicted by the burst outflow model 
in the bottom panel of Figure 8. As shown in Figure 5, such models eject a large amount 
of mass into the CGM within $\sim 300-400$ Myr, and in 1-2 Gyr the total mass 
decreases by more than $50\%$ due to heat evaporation or gas recycling. 
The color lines show the evolution of covering fraction as a function of time. 
As can be seen, the covering fraction in the inner region decreases from 
$100\%$ to $10\%$ over the period from 400 Myr to 1600 Myr. This demonstrates that 
to maintain the amount of mass in the CGM over a long period time, a replenishment 
of the circumgalactic gas is required.

\subsubsection{Inflow models}

We now investigate the spatial distribution of gas clouds from the inflow models. 
Expected to occur around both star-forming and passive galaxies, gas inflows are a possible source of the cool gas observed around passive galaxies. In the following, we explore such a scenario by modeling the spatial distribution of cool gas clouds falling from the outskirt of the halos and compare the output with observations.
The model assumes that gas clouds are initially located 
at 200 kpc from the halo center and fall towards the center of the halo with 
a constant mass flow rate of $20 \, \rm M_{\odot}/yr$. The system is evolved over 4 Gyr. 
As shown in Figure 6, the amount of mass in CGM becomes stable after about 
2 Gyr. The upper panel of Figure 10 shows the average covering fraction of the 
inflow gas clouds over a period from 2000 Myr to 4000 Myr for sightlines of 
different $N_{\rm HI}$. Results are shown for different cloud mass functions:
power laws with $\alpha=0$ and $-1$ in purple and orange, respectively, 
and a Schechter function in green. The red and blue data points show the observed covering 
fraction based on MgII absorbers around passive and star-forming galaxies from \citet{Lan2014}.

As in the outflow models, the covering fraction of the inflowing gas as a 
function of the HI column density depends on the shape of the initial 
mass function. The power law model with $\alpha=0$ (purple) produces only massive clouds, while steeper initial mass functions ($\alpha=-1$, orange, and a Schechter like, green) can produce more small clouds and 
therefore larger gas covering fractions that match better the observed ones.

Let us have a close look at the dependence of the shape of the covering fraction 
on $\rm N_{HI}$ predicted by the Schechter model of initial cloud mass 
function (green lines). The covering fraction of $\rm N_{HI}>10^{19.6} \, cm^{-2}$ 
(right panel) increases monotonically towards the center, while the covering 
fraction of $\rm N_{HI}<10^{19.6} \, cm^{-2}$ (left and middle panels) 
becomes flat or even decreases slightly towards the center. This difference is 
due to projection effect. The monotonic increasing trend of covering fraction 
for $\rm N_{HI}>10^{19.6} \, cm^{-2}$ is produced by relatively massive clouds 
that can survive long time and reach to the inner region of the halo where the 
clouds have higher $\rm N_{HI}$, as shown in Figure 2. In contrast, due to 
the evaporation, there is no small clouds that can survive long enough to reach 
to the center and contribute to the covering fraction of $\rm N_{HI}<10^{19.6} \, cm^{-2}$. 
Thus, the covering fraction of $\rm N_{HI}<10^{19.6} \, cm^{-2}$ within $r_{p}<100$ kpc 
is due to clouds at large distances $>100$ kpc but projected in 
the plane perpendicular to the line of sight. The projection yields to a flat 
covering fraction in the inner region. We note that for all the cases  
we have explored, the modeled covering fraction for clouds with 
$\rm 10^{18.6}<N_{HI}<10^{19.1} \, cm^{-2}$ is always lower than the observed 
ones as shown in Figure 10. This is due to the fact that, with a given mass, 
only within a narrow range of locations beyond 100 kpc 
that clouds can have $\rm 10^{18.6}<N_{HI}<10^{19.1} \, cm^{-2}$, 
as shown in Figure 2. These results show that the inflow model can reproduce 
the observed covering fractions of high $\rm N_{HI}$ systems around passive galaxies 
within 100 kpc, but it fails to reproduce the low $\rm N_{HI}$ systems.

\subsubsection{In-situ models}

Gas clouds that form in-situ in the halos due to thermal instability and cooling \citep[e.g.,][]{Kaufmann2006, Hummels2018} are another possible source for the cool gas observed around passive galaxies. We now consider and model this scenario by  populating clouds falling in the halos with initial positions following a NFW and a power law, $n(r)\propto r^{-1}$, density distributions. The cloud mass formation rate is $20 \, \rm M_{\odot}/yr$ with the Schetchter function of 
$\alpha=-1$ and $M_{*}=10^{5.5} M_{\odot}$. The covering fraction of clouds from the NFW profile (orange) 
and the power law (purple) are shown in the lower panel of Figure 10. 

The surface mass density 
profile corresponding to a NFW profile is about $\Sigma \propto r_{p}^{-1}$, similar to the 
profile of the observed covering fraction. However, with a given cloud mass, the corresponding 
$\rm N_{HI}$ decreases with $r_{p}$ because of the hot gas profile shown in Figure 2. These two 
factors together produce a sharp decrease of the covering fraction beyond 100 kpc.

In comparison to the NFW profile, the power law distribution produces a constant surface 
density profile,  with more mass located in the outskirt.  
Consequently, the profile of the covering fraction is more extended, as shown in the lower panel of Figure 10.  
However, the covering fraction in the inner region is lower than that predicted by the NFW profile. 
The inner profile of the covering fraction is also flatter due to the fact 
that the covering fraction is mostly contributed by clouds at larger distance in 
three-dimensional space. The power law distribution predicts covering fractions that are similar to the 
inflow models, but with an enhancement of covering fraction for high $\rm N_{HI}$ systems by a factor of $\sim 1.5-2$ in the inner region due to 
clouds directly formed in-situ close to the center. 

Comparing with the observed covering fractions, we conclude that both the gas clouds falling from the outskirt of the halo and clouds directly formed in-situ are able to explain the observed 
high $\rm N_{HI}$ systems around passive galaxies, but both fail to explain the 
low $\rm N_{HI}$ systems.

%======================================================
%Figure 
%======================================================
\begin{figure}[t]
\center
\includegraphics[scale=0.4]{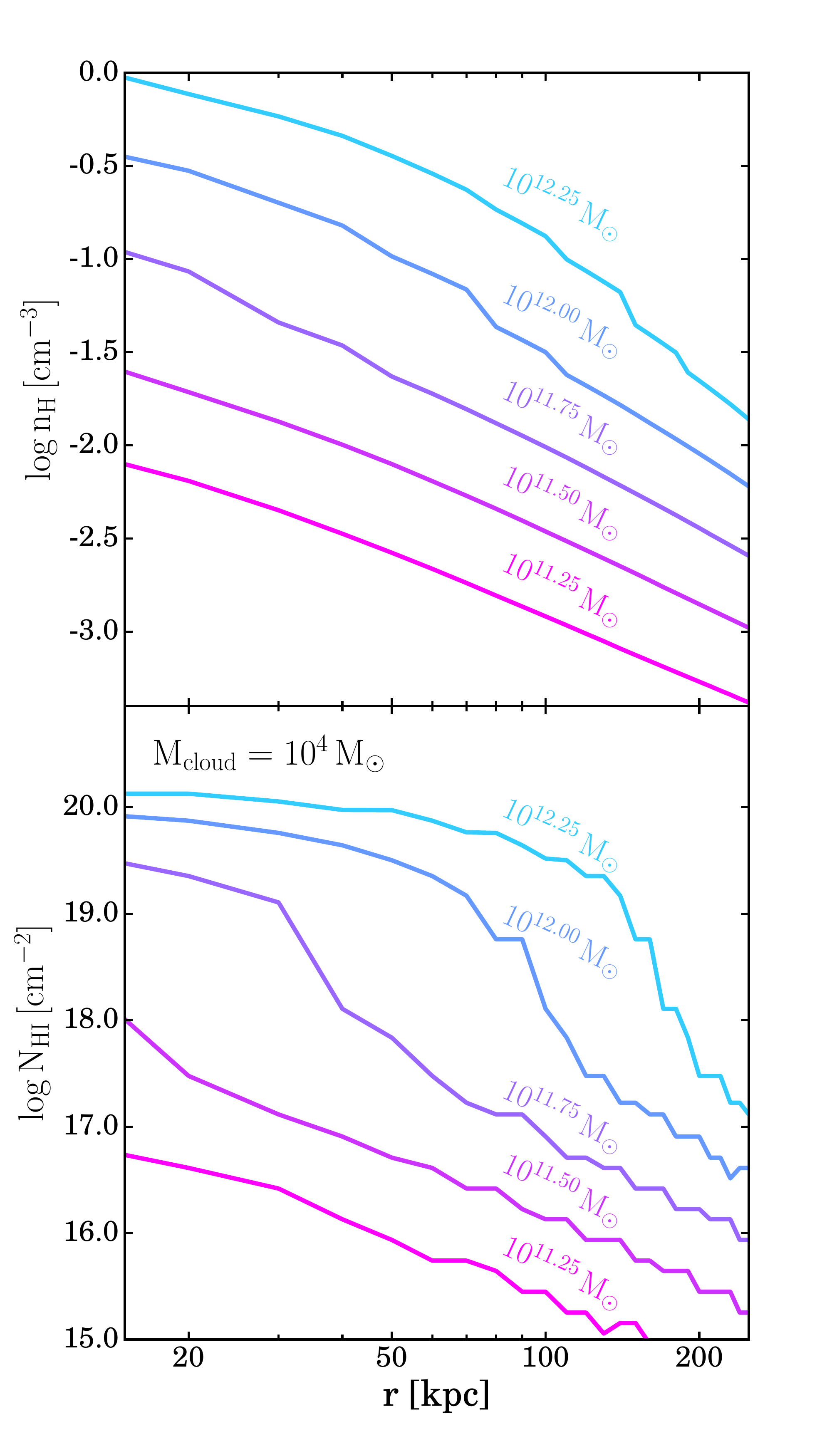}
\caption{Hydrogen volume density and neutral hydrogen column density as a function of halo mass and position. \emph{Top:} Hydrogen volume density. \emph{Bottom:} Neutral hydrogen column density for clouds with $10^{4} \, M_{\odot}$.}
\label{fig:halo_mass_dependence}
\end{figure}
%======================================================

%======================================================
%Figure 
%======================================================
\begin{figure}[t]
\center
\includegraphics[scale=0.6]{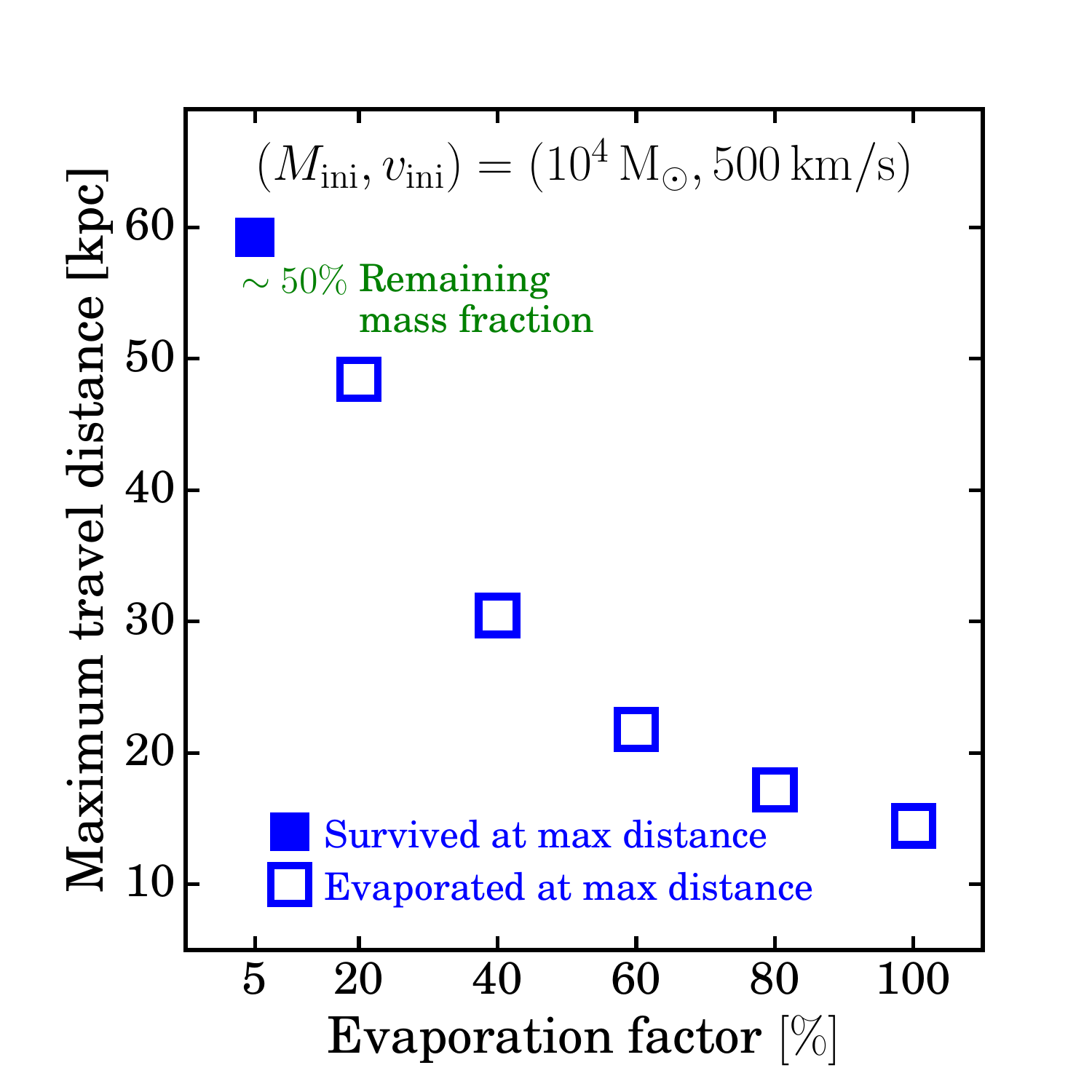}
\caption{The effect of the evaporation factor on the maximum travel distance of clouds. With 5\% efficiency, clouds with 500 km/s and $\rm 10^{4}\, M_{\odot}$ can reach 60 kpc with $\sim$50\% of initial mass, while with 100\% efficiency, the clouds lose all the mass at 15 kpc.}
\label{fig:f_distance_dependence}
\end{figure}

\section{Discussion}

\subsection{Dependence on other model parameters}

Here we discuss how the properties of the cool gas depend on other parameters in the model.

\subsubsection{Dependence on hot gas}

The properties of the cool clouds are affected by the adopted model for hot gaseous 
halos.  Since the hot gas properties correlate with the halo masses in our model, we illustrate such 
an effect by calculating the properties of the cool gas as a function of halo mass while keep other 
parameters intact. Figure~\ref{fig:halo_mass_dependence} shows the dependence of the hydrogen volume 
density and the neutral hydrogen column density on the halo mass. For the hot gas model adopted here, 
the pressure of hot gas at a given radius scale with the halo mass as 
$\rm M_{\rm halo}^{2}$, as reflected in the volume density of the cool gas shown in the upper panel. 
The lower panel shows the neutral hydrogen column density of clouds with a fixed cloud mass 
$10^{4} \, M_{\odot}$ as a function of halo mass indicated by the color lines. We find that clouds 
with $\rm N_{HI}>10^{19} \, cm^{-2}$ can only be found around halos with masses greater than 
$\sim 10^{11.75} \, M_{\odot}$ due to the fact that hydrogen in cool clouds becomes highly 
ionized when the volume density is low. 

This result suggests that absorbers with high column density in neutral hydrogen e.g. 
$N_{\rm HI}>10^{19} \rm cm^{-2}$ and single-ionized gas (e.g., MgII, SiII) only exist in halos 
that can provide sufficient pressure-support from hot ambient gas. In addition, the high column density 
systems are preferentially found near the center of the halos where the pressure of the hot gas is 
expected to be higher. Thus, this model of pressure-confined gas cloud predicts that the incidence 
rate of absorbers with high column density in neutral hydrogen and in 
single-ionized metal elements increases with halo mass while decreases with radius. 
This trend is consistent with observations that high neutral hydrogen column density 
systems, such as strong MgII absorbers, sub-DLAs and DLAs,  tend to be found around massive 
galaxies \citep[e.g.,][]{Chen2010, Lan2014, Kanekar2018}.

\subsubsection{Evaporation factor} 
In Equation (8), we introduce a factor $f$ to control the efficiency of cloud mass loss due to 
evaporation. This factor affects the amount of mass in 
the circumgalactic medium by reducing the lifetime of clouds. In our model, we set $f$ to 
be $5\%$. If $f$ is taken to be $100\%$, i.e. 20 times as high as we adopt, 
the gas flow rate needs to be increased by a factor of 20 in order to maintain the same 
amount of the mass in the CGM. This leads to an unreasonably high value of the gas flow rate,
about 400 $\rm M_{\odot}/yr$ for the constant outflow/inflow models
to match the observed covering fractions. In addition, high $f$ will reduce the travel 
distances of outflow clouds and, therefore, the extension of the cool circumgalactic clouds 
originated from galaxies. 

In Figure~\ref{fig:f_distance_dependence}, we show the effect of the evaporation efficiency 
on the travel distance of gas clouds ejected from the center with 500 km/s initial velocity 
and $10^{4} \, M_{\odot}$ initial cloud mass. The figure shows that with the evaporation factor 
we adopt, 5\%, the gas cloud can travel to about 60 kpc from the center while retain about 50\% 
of the initial mass at the maximum distance determined by gravity and the ram pressure of the system. 
In contrast, for higher evaporation factors, gas clouds lose all the mass while moving outwards 
and the maximum distances are set by the evaporation. With 100\% evaporation factor, the 
maximum travel distance is only 15 kpc. This result illustrates that low evaporation 
factor is needed to produce the cool CGM with reasonable parameter values.

\subsection{The evaporated gas in the CGM} 

Our model shows that a significant fraction ($\sim50\%$ for $\alpha=-1$ and $\sim 80\%$ for a 
Schechter mass function) of total ejected cloud mass is evaporated. This suggests that the 
circumgalactic medium also contains the evaporated gas. 
Although the physics of the evaporated gas is not well-developed and not included in our model,
the temperature and the density of the evaporated gas 
are expected to be between that of pressure-confined gas clouds and that of the hot ambient 
gas over a certain timescale \citep[e.g.,][]{Balbus2016}. This may be investigated with high 
resolution hydrodynamic simulations for individual clouds 
\citep[e.g.,][]{Armillotta2017, Liang2018, Sparre2018}. For example, \citet{Armillotta2017} 
demonstrate that the interaction between the hot ambient gas and the pressure-confined clouds 
can produce evaporated and stripped gas with relatively low column density, and highly ionized species, such as SiIII and OVI (their Figure 3 and Figure 6). These results demonstrate that the evaporated gas may be responsible 
for a significant fraction of the relatively highly-ionized absorption line systems, 
such as CIV and OVI, and perhaps some low column density systems of lowly-ionized 
species, such as MgII. It is important to understand the properties 
of such gas in detail in order to have a complete picture of the CGM.

\subsection{Comparisons with other studies}

\textbf{A universal density model} -
Motivated by observations, \citet{Stern2016} develop a hierarchical density structure model 
for the circumgalactic medium assuming that the density structure of the circumgalactic medium 
follows a power law distribution with high density gas embedded in the low density one. 
They show that this density structure can simultaneously reproduce the column densities of metal 
species in neutral (e.g., MgI) to highly ionized phase (e.g., OVI) at redshift 0.1. Despite that this model and our model start with different assumptions, the two predict similar MgII 
gas properties: cloud sizes about 50 pc and cloud masses about 
$\rm 10^{2}-10^{3}\, M_{\odot}$ (See their Table 1). 

\textbf{Shattering cloudlet model} - 
Recently, \citet{McCourt2016} propose that the pressure-confined cool gas clouds will be fragmented into 
tiny droplets with a characteristic scale with $\rm N_{H}\sim10^{17} cm^{-2}$ corresponding to the 
scale of the product of sound velocity and cooling time. This property has been demonstrated in 
high-resolution hydrodynamical simulations \citep[e.g.,][]{McCourt2016, Liang2018, Sparre2018}. 
As discussed in \citet{McCourt2016}, for a single droplet with $\rm N_{H}\sim10^{17} \, cm^{-2}$, 
the gas is optical thin and highly ionized; However, a collection of droplets can be self-shielded 
and produce systems with high neutral hydrogen column densities. In some sense, a single cloud in our 
model can be considered as a collection of droplets which are physically associated with each other 
and move together.

\textbf{Cool gas inconsistent with the pressure equilibrium scenario} - 
Using the data from COS on the Hubble telescope, \citet{Werk2014} characterized the 
physical properties of the gas around galaxies at redshift$\sim0.1$. Using the CLOUDY simulation, they 
found that the detected cool gas has volume density to be $10^{-3}-10^{-4} \, \rm cm^{-3}$, 
inconsistent with the density of pressure-confined cool gas in $10^{12}\, \rm M_{\odot}$ halos. 
They concluded that the scenario in which cool circumgalactic gas is in the pressure 
equilibrium is a poor description of the CGM. However, as shown in this study, the gas clouds 
in pressure equilibrium are expected to be most neutral with high neutral hydrogen column densities 
$\rm N_{HI}>10^{18.5} \, cm^{-2}$, and with the covering fraction traced by strong MgII absorbers 
at most $20-30\%$ at 50 kpc and $5-10\%$ at 100 kpc. Having 44 sightlines intercepting the CGM of 
galaxies with a wide range of halo masses from $10^{11.3}$ to $10^{13.5} \, M_{\odot}$ 
\citep{Tumlinson2013}, it is possible that the pressure-confined gas clouds are not well sampled 
by the observation and most of the sightlines intercept some diffuse gas, such as 
evaporated gas in our model, which may have a range of temperatures and may not be 
in pressure equilibrium with the hot gas. 
In addition, it is possible that a single phase model is not adequate to capture the underlying multiphase gas structure as illustrated in \citet{Stern2016}.
Given above, we argue that more data and advanced 
modeling are required to better test the scenario of the pressure equilibrium.

\subsection{Limitations and directions for future work}

\textbf{Hot gas} - As shown in this work, the pressure confined cool gas properties depend on the 
assumed hot gas properties. In our model, we assume that the hot gas is in hydrostatic equilibrium. 
However, it may not reflect the real properties of hot gas as suggested by \citet{Oppen2018}. 
In addition, we only consider the hot gas profile from a halo with $10^{12}\, M_{\odot}$ when comparing with observations. However, the observed covering fractions are expected to be obtained from a galaxy population living in halos with a range of mass. For future studies, one needs to 
take this halo mass dependence into account.

\textbf{Cool gas} - we calculate the physical properties of the cool gas by assuming that the 
cloud temperature is a constant $10^{4}$ K. However, it is expected that the cloud temperature 
can vary depending on the heating mechanisms. We have performed a calculation by adjusting 
the cloud temperature. If the cloud temperature is assumed to be 5000 K, the column density, $\rm N_{HI}$, 
will be enhanced for a given cloud mass. Therefore, for the initial mass function with 
$\alpha=-1$ and $\rm M*=10^{5.5} \, M_{\odot}$ adopted in this work, the model will produce more 
$\rm N_{HI}>10^{19.5} \, cm^{-2}$ systems and overproduce the gas covering fraction for such systems.
On the other hand, by increasing the temperature of clouds to be 20000 K, all the gas clouds become highly 
ionized with $\rm N_{HI}<10^{19} \, cm^{-2}$. Such a scenario cannot produce any high column 
density systems as observed. This suggests that the adopted cloud temperature with $10^{4}$ K 
is perhaps reasonable for reproducing the gas properties of high $\rm N_{HI}$ gas absorbers. 

\textbf{Survival of clouds} - 
One of the key assumptions in our model is that the clouds are not destroyed by the 
hydrodynamic instability. Although a number of studies have shown that various mechanisms can suppress 
the effect of hydrodynamical instability \citep[e.g.,][]{Vietri1997, McCourt2015, Armillotta2017, Gronke2018}, 
it is still unclear how these mechanisms operate in reality. This problem is related to the problem 
of entraining the cool gas along with a hot wind \citep[e.g.,][]{Zhang2017}. However, we argue that 
the existence of cool gas and the evidence of contribution from outflows in the halos 
\citep[e.g.,][]{Bordoloi2011, Lan2018} suggest that either the cool gas outflows must survive 
over the hydrodynamic instability, as we assume in our model,
or cool gas is redistributed in the halos by some other mechanisms 
associated with outflows \citep[e.g.,][]{Thompson2016,Lochnaas2018}.

\textbf{Gas properties at large scales - }
Our model has difficulties in explaining systems  
with $\rm N_{HI}>10^{19} \, cm^{-2}$ beyond 100 kpc 
\citep[e.g.,][]{Zhu2014,Huang2016,Lan2018} 
due to the low volume density of the hot ambient gas as shown in Figure 2. 
It is, therefore, 
likely that some of the systems at large distances are produced by neighboring 
halos due to spatial clustering of halos. Such `two-halo' term is not included 
in our current model, but should be modeled.

\section{Summary}
We develop a flexible semi-analytic framework to explore the physical properties of the cool 
pressure-confined circumgalactic clouds with mass from 10 $M_{\odot}$ to $10^{8} \, M_{\odot}$. 
We take into account the effects of gravity and interaction between hot gas and cool gas 
and model the trajectory, the lifetime, and the observed properties of the cool clouds with 
CLOUDY simulations. The ensemble properties of the cool CGM are explored by populating clouds 
following various mass functions in the halos with three origins, outflows, inflows, and 
in-situ formation. With this framework, we investigate how different mechanisms affect the 
observed properties of the cool CGM. Our results are summarized as follows:
\begin{enumerate}
    \item The pressure-confined cool gas clouds have $\rm N_{HI}>10^{18.5} \, cm^{-2}$ with sizes 
    about 10-100 pc similar to the observed properties of strong metal absorbers, sub-DLAs, and DLAs 
    observed towards background spectra. 
    
\item We illustrate the effect of hot gas on the motion and lifetime 
    of cool gas clouds; the ram pressure from the hot gas prevents most of the high velocity cool clouds 
    from escaping the system and heat evaporation destroys small clouds in about hundred million years. Because of these mechanisms and gas recycling, the cool CGM is transient and therefore a constant replenishment is required to sustain it.
  
\item We demonstrate that the morphology and structure of the CGM depend on the shape of the initial cloud mass function. A Schechter-like mass function is favored in order to produce covering fractions as a function of HI column density that are comparable to the observed ones.
    
\item We show that only with certain combinations of parameters can the model reproduce the 
three gas properties of star-forming galaxies: the spatial distribution, down-the-barrel outflow absorption, and gas velocity dispersion at the same time. This demonstrates the constraining power of the observed CGM properties on physical models of outflows.
    
\item We show that high $\rm N_{HI}$ gas around passive galaxies may also be 
explained by in-situ gas clouds and gas inflow, while such models under-produce low $\rm N_{HI}$ systems.

\item We demonstrate how the cool gas properties depend on the assumed halo mass and 
    evaporation efficiency. This model predicts that high neutral hydrogen 
    column density systems should tend to be found in massive halos and close to the center where 
    the hot gas pressure is expected to be the highest. In addition, to maintain the cool CGM with a 
    more realistic gas flow rate, the heat evaporation factor needs to be much lower than the classical efficiency.  
\end{enumerate}
These results demonstrate that analytic and semi-analytic approaches are another promising way to 
understand and explore physical mechanisms that govern the properties of the cool CGM without 
limited by the numerical resolution as indicated by recent simulations and theoretical 
models \citep[e.g.,][]{McCourt2016, Sparre2018,vandevoort2018,Peeples2018,Hummels2018}. 
Finally, we list two key properties of the pressure-confined cool CGM which can be tested by future observations:
\begin{itemize}
    \item The pressure of hot gas halo increases with halo mass and decreases with impact parameter. These dependences are expected to be reflected in the properties of the pressure-confined cool CGM. With a statistical sample, one can possibly map out the pressure profile of cool gas as a function of halo mass and impact parameter and confront the observation with this model \citep[e.g.][]{Werk2014,Zahedy2018}. 
    
    \item Based on the blueshifted absorption line profile observed down-the-barrel, the typical velocity of cool gas outflows is about a few hundred km/s. Our model shows that most of cool gas clouds with such initial velocities can not escape the halos. They are mostly confined within 100 kpc (Figure 3). Therefore, the gas clouds are expected to enhance the covering fraction and metallicity within such a scale. This scenario can be tested by probing the covering fraction \citep[e.g.][]{Lan2018} and metallicity profiles \citep[e.g.][]{Peroux2016} of cool gas around star-forming galaxies as a function of impact parameters. 
\end{itemize}

\acknowledgements
We thank J. Xavier Prochaska, Neal Katz, and Brice~M\'enard for insightful discussions.  We also want to thank the anonymous referee for the constructive report.
HJM acknowledges support from NSF AST-1517528,  and from 
National Science Foundation of China (grant Nos. 11673015, 11733004). Kavli IPMU is supported by World Premier International Research Center Initiative of the Ministry of Education, Japan.

\end{document}